\documentclass{aa}
\usepackage{graphics}

\begin{document}

\thesaurus{06 (08.02.6; 08.05.1; 08.16.5; 08.09.2 HR\,4796; 13.25.5)}

\title{X-ray emission from Lindroos binary systems}

\author{N. Hu\'elamo\inst{1} 
\and R. Neuh\"auser\inst{1} 
\and B. Stelzer\inst{1}
\and R. Supper\inst{1}\and H. Zinnecker\inst{2}}

\institute{Max-Planck-Institut f\"ur extraterrestrische Physik, 
 D-85740 Garching, Germany
\and Astrophysikalisches Institut Potsdam, An der Sternwarte 
 16, D-14482 Potsdam, Germany\\} 

\offprints{N. Hu\'elamo}
\mail{huelamo@xray.mpe.mpg.de}
\date{Received/Accepted}

\maketitle

\begin{abstract}

   We present a study of the X-ray emission from binary systems
   extracted from the Lindroos catalogue (Lindroos 1986) based on
   the ROSAT All-Sky survey as well as  ROSAT PSPC and HRI pointings.  The
   studied sample consists of visual binary systems 
   comprised of early-type primaries and late-type secondaries.  
   The ages of the systems were determined by Lindroos (1985) from
   uvby$\beta$ photometry of the primaries. These ages range between 33 
   and 135 Myr, so if the late-type secondaries are physically bound to 
   the early-type primaries, they could be  Post-T Tauri stars (PTTS).  

   We have found strong X-ray emission from several secondaries. 
   This fact together with their optical and IR data, 
   make them {\it bona fide} PTTS candidates.
   We have also detected X-ray emission from several
   early-type primaries and, in particular, from most of the 
   late-B type stars. Because their HRI hardness ratios are
   similar to those from resolved late-type stars, 
   the presence of an unresolved
   late-type companion seems to be the cause of this emission.
      

\end{abstract}

\section{Introduction}

   Pre-main sequence (PMS) late-type stars are known to be X-ray
   sources (see Walter et al. 1988, Bouvier 1990 and Neuh\"auser et al. 1995).
   In the evolution of these stars to the Main Sequence (MS), 
   there is a state usually defined as Post-T Tauri Stars (PTTS). 
   PTTS were first defined by Herbig (1978) as PMS stars 
   more evolved than Classical T Tauri stars (CTTS) 
   but still contracting to the MS. 
   Given that the stage of CTTS is only a small fraction of the total 
   time of contraction of low-mass stars to the MS, 
   PTTS should be much more abundant than CTTS
   if star formation has been ongoing for a sufficiently long time.
   However, it is difficult to find PTTS because 
   they do not show spectroscopic or photometric peculiarities 
   which make them easy to detect. Unlike CTTS, 
   they do not present significant IR or UV excesses 
   and the H$\alpha$ line is not observed as a strong emission line. 
   Therefore, the identification  of these 
   stars is difficult and relies on the detection of the Li {\sc I} (6708\AA) 
   absorption line and the chromospheric Ca {\sc II} (H \& K) lines 
   in their spectra, as well as on their X-ray detection.  
    
    Murphy (1969) first proposed that PTTS could be searched as
   members of young binary systems. The basis of his idea was that 
   the MS lifetime of high-mass stars is comparable to the contraction 
   timescale of solar-type stars. Hence, there could be binary systems 
   comprised of MS early-type stars physically bound to PTTS.
   Gahm et al. (1983) carried out photometric and spectroscopic
   observations of visual double stars with early type primaries.
   The derived data, together with $JHKL$ observations, allowed 
   Lindroos (1986; L86 hereafter) to identify 78 likely physical 
   pairs with several PTTS candidates as secondaries. 
   
    A high lithium abundance and a high chromospheric activity 
   level are necessary (although not sufficient) indicators of  youth. 
   Mart\'{\i}n et al. (1992) and Pallavicini et al. (1992)
   carried out optical spectroscopy of the Lindroos late-type companions,
   detecting the Li {\sc I}(6708\AA) absorption line 
   and the Ca {\sc II} (H \& K) emission lines in the spectra of 
   several PTTS candidates. 
   Ray et al. (1995) took this sample of ``genuine'' PTTS and looked for
   circumstellar matter around them. For this purpose, they
   analyzed the IRAS database (Point Source Catalogue and Faint 
   Source Catalogue) and also searched for continuum 1.1 mm emission. 
   While IR excesses were found for most of the sources, 
   no mm dust continuum was detected (see also Gahm et al. 1994 
   and Jewitt 1994).  
    	
   \begin{table*}[htbp]
      \caption{Stellar Data of the binary sample}
       \label{GenDat}
 	\begin{flushleft}
        \begin{tabular}{llllrrllll}
          \noalign{\smallskip}
          \hline
          \noalign{\smallskip}

HD  & Sp.Type$^1$ & \multicolumn{3}{c}{Optical position$^2$} &  Sep.$^1$ & 
  Distance$^3$ & V$_{A}$, V$_{B}$ $^4$ & A$_v$$^5$ & Clas.$^6$\\
\cline{3-5}
    &   & RA (2000) & & Dec (2000) &  (\arcsec) & (pc)       & (mag) & (mag) \\
    &   & (h) (m)  (s) & & ( \degr) (\arcmin) (\arcsec) & & & & \\
          \noalign{\smallskip} 
          \hline 
          \noalign{\smallskip}
560  A+B   & B9V+G5Ve  & 00 10 02.20 & & 11  08 44.93 & 7.7 & 100$\pm$9 & 5.53,10.37 & 0.00  & CP\\
1438  A+B  & B8V+F3V   & 00 18 42.17 & & 43  47 28.11 &  6.2 & 212$\pm$35 & 6.11,9.7$^* $ & 0.04 & \\
8803   A+B & B9V+F6Vp  & 01 26 53.55 & & 03  32 08.32 &  6.0 & 160$\pm$25 & 6.43,9.67  & 0.12 & LO\\
17543  A+C & B6IV+F8V  & 02 49 17.56 & & 17  27 51.51 & 25.2 & 185$\pm$37 & 5.28,10.73 &  0.21& PP \\ 
23793  A+B & B3V+F3Vp  & 03 48 16.27 & & 11  08 35.86 &  9.0 & 173$\pm$31 & 5.10,9.41 & 0.08 & LO \\
27638  A+B & B9V+G2V   & 04 22 34.94 & & 25  37 45.53 & 19.4 & 82$\pm$8  & 5.41,8.43  & 0.00 & PP\\
33802  A+B & B8V+G8Ve  & 05 12 17.90 & & -11 52 09.19 & 12.7 & 74$\pm$4  & 4.47,9.92  &  0.01 & CP \\ 
35007  A+C & B3V+G3V   & 05 21 31.84 & & -00 24 59.36 & 37.6 & 330$\pm$89 & 5.68,11.88   &0.16 & \\
36013  A+B & B2V+F9V   & 05 28 45.28 & &  01 38 38.17 & 25.0 & 303$\pm$86 & 6.89,12.49  & 0.04 & \\
36151  A+B & B5V+G1V   & 05 29 25.4  & & -07 15 39.18 & 48.6 & 370$\pm$122 & 6.69,10.60    & 0.08 & \\
36151  A+X & B5V+G7V   & 05 29 25.4  & & -07 15 39.18 & 45.0 & 370$\pm$122  & 6.69,11.98   & 0.08 & \\
36779  A+B & B2.5V+K5IV & 05 34 03.89 & & -01 02 08.61 & 27.5  & 380$\pm$378: & 6.24,11.20 & 0.10 & LO \\
38622  A+C & B2V+G2V   & 05 47 42.91 & & 13  53 58.56 & 24.9 & 245$\pm$54 & 5.27,12.01  & 0.04 & CP\\ 
40494  A+B & B3IV+G8V  & 05 57 32.21 & & -35 16 59.80 & 33.8 & 263$\pm$38 & 4.36,12.66 &  0.00 & PP\\ 
43286  A+B & B5IV+G3V  & 06 15 30.24 & & 03  57 29.50 & 18.3 & 308$\pm$85 & 6.99,12.38  & 0.06 & \\
48425  A+C & B3V+G5V   & 06 42 27.57 & & -23 13 57.29 & 34.9 & 368$\pm$122 & 6.9,10.4$^*$ &  0.00 & \\
53191  A+B & A0V+G3V    & 07 00 16.73 & & -60 51 45.83 & 17.0 & 207$\pm$25 & 7.74,11.75& 0.05 & PP \\ 
53755  A+B & B0.5IV+F5III & 07 05 49.64 & & -10 39 36.28 & 6.2  & 1087: & 6.49,10.3$^*$ & 0.65 & \\
56504  A+B & B9IV+G6III & 07 16 00.27 & & -29 29 33.81 & 31.1 & 932$^+$ & 9.8,9.7$^*$ &  0.07 & \\
60102  A+B & B9.5V+G8V  & 07 11 38.41 & & -84 28 09.77 & 16.4 & 206$\pm$25 & 7.54,11.86 & 0.22 & PP\\ 
63465  A+B & B2.5III+F2V& 07 47 24.99 & & -38 30 40.12 & 10.9 &  380$\pm$80 & 5.08,11.13 & 0.25 & LO\\
70309  A+B & B3IV+K2IV  & 08 19 05.58 & & -48 11 52.27 & 42.5 & 252$\pm$36 & 6.45,11.28  & 0.12 & LO\\
71510  A+C & B3V+G3V    & 08 25 31.32 & & -51 43 38.69  & 34.9  & 207$\pm$21 & 5.19,10.77& 0.01 & LO\\
74146  A+B & B5V+F0IV   & 08 39 57.59 & &-53 03 17.03 & 16.6 & 131$\pm$8 & 5.19,8.66$^*$ & 0.03 & \\ 
76566  A+B & B3V+G4V    & 08 55 19.20 & &-45 02 30.01 & 35.0 & 286$\pm$51 & 6.28,12.64 & 0.00 & LO\\
77484  A+B & B9.5V+G5V  & 09 02 50.65 & & 00 24 29.54 & 4.4  & 250$\pm$64 & 8.02,12.0   & 0.09 & PP \\
86388  A+B & B9V+F5V    & 09 55 05.60 & &-69 11 20.31 & 9.2  & 177$\pm$21 & 6.87,9.98 & 0.02 & LO \\ 
87901  A+B & B8V+K0Ve   & 10 08 22.31 & & 11 58 01.94 & 176.9& 23.7$\pm$0.4 & 1.35,8.08 & 0.00 & LO \\
90972  A+B & B9.5V+F9Ve & 10 29 35.38 & &-30 36 25.43 & 11.0 & 147$\pm$16 & 5.58,9.65 & 0.03  & CP \\ 
104901 A+B & B9II+F0IIe & 12 04 46.98 & &-61 59 48.60 & 23.0 & 980: & 7.43,8.01$^*$  & 1.11 & \\
106983 A+B & B2.5V+ G8III & 12 18 26.24 & & -64 00 11.05& 33.8 & 110$\pm$6 & 4.05,12.49 & & LO \\
108767 A+B & B9.5V+K2Ve & 12 29 51.85 & & -16 30 55.56 & 24.2 & 27$\pm$9  & 2.94,8.43 &  0.00 & CP \\ 
109573 A+B & A0+M2.5    & 12 36 01.3  & & -39 52 09.00  & 7.6  & 67$\pm$3  & 5.78,13.3 &  0.00 &   \\ 
112244 A+B & O9Iab+K0III& 12 55 57.13 & & -56 50 08.90 & 29.1 & 578$\pm$213 & 5.38,11.77& 0.90 & LO \\
112413 A+B & A0IIIp+F0V & 12 56 01.67 & &  38 19 06.17 & 19.6 & 34$\pm$1  & 2.90,5.60$^*$ & 0.00 &  \\
113703 A+B & B4V+K0Ve   & 13 06 16.70 & & -48 27 47.84 & 11.4 & 127$\pm$12 & 4.72,10.8 &  0.00 & CP\\
113791 A+B & B2III+F7V  & 13 06 54.64 & & -49 54 22.49 & 25.1 & 126$\pm$13 & 4.25,9.38 &   & CP \\
120641 A+B & B8V+F0Vp   & 13 52 02.98 & & -52 48 05.81 & 18.0 & 333: & 7.47$^*$,-$^7$ &  0.00 & \\
123445 A+B & B9V+K2V    & 14 08 51.89 & & -43 28 14.80 & 28.6 & 218$\pm$37 & 6.19,12.52 & 0.12 & LO \\
127304 A+B & A0V+K1V    & 14 29 49.67 & &  31 47 28.28 & 25.8 & 106$\pm$8 & 6.07,11.37 & 0.04 & PP\\
127971 A+B & B7V+K0V    & 14 35 31.48 & & -41 31 02.77 & 26.9 & 109$\pm$9 & 5.89,11.22 & 0.08 & LO\\
129791 A+B & B9.5V+K5Ve & 14 45 56.2  & & -44 52 03.08 & 35.3 & 129$\pm$16 & 6.94,12.93 & 0.26 & CP \\
137387 A+B & B3IVe+K5IV & 15 31 30.82 & & -73 23 22.53 & 27.0 & 312$\pm$59 & 5.47,11.27 & 0.47 & LO \\
138800 A+X & B8IV+KOV   & 15 40 21.33 & & -73 26 48.07 & 34.0 & 225$\pm$41 & 5.65,12.86  & 0.22 & LO\\
143939 A+B & B9III+K3Ve & 16 04 44.49 &  &-39 26 04.76 & 8.6  & 167$\pm$27 & 6.98,11.80& 0.00 & CP \\
145483 A+B & B9V+F3V    & 16 12 16.04 &  &-28 25 02.29 & 4.6  & 91$\pm$8  & 5.67$^*$,-$^7$ &0.25 & \\ 
162082 A+B &  B7V+F2V   & 17 50 18.10 &  &-26 19 33.87 & 10.7 & 314$^+$ & 8.16,11.05 &  0.63 & LO\\
174585 A+B & B2.5V+K2IV & 18 49 45.91 &  & 32 48 46.15 & 34.8 & 310$\pm$52 & 5.90,10.89$^*$  &  0.17 & \\
174585 A+C & B2.5V+G0V  & 18 49 45.91 &  & 32 48 46.15 & 58.7 & 310$\pm$52 & 5.90,10.3$^*$  &  0.17& \\
180183 A+B & B3V+K05    & 19 18 41.54 &  &-56 08 40.90 & 19.4 & 244$\pm$49 & 6.82,11.45  & 0.10 & LO\\
        \noalign{\smallskip}
        \hline \\
        \end{tabular}
       \end{flushleft}  
{\bf Notes:}
1. Taken from Pallavicini et al. (1992); 2. Coordinates of the primary star; 
3. Deduced from Hipparcos parallax of the primary star,
except those marked with $^+$ taken from L86; 4. Adopted from Pallavicini 
et al. (1992) except those marked with an asterisk
obtained from the SIMBAD  database; 5. taken from L86; 
6. Classification taken by Pallavicini 
et al. (1992) with respect to the presence of the Lithium 
absorption line and CaII (H and K) chromospheric lines in the 
spectrum of the secondary: CP (certainly physical), PP 
(probably physical) and LO (likely optical); 7. Unknown visual 
magnitude for the secondary star.

   \end{table*}
   
   The X-ray emission from Lindroos binary systems was first  
   studied by Schmitt et al. (1993). After the analysis of 
   seven pairs comprised of late B-type stars and 
   later-type companions, the main result was the detection of 
   X-rays from both members of the pair. In the case of late-type 
   stars it is well-known that they produce X-rays 
   in their hot coronae. However, this is not the case of late-B 
   type stars. Theoretically, early-type stars between B4 and A7 are not
   expected to be X-ray emitters: they do not possess the strong winds 
   thought to be responsible of the X-ray emission in O- and early B-type 
   stars (Lucy \& White 1980), nor significant convection 
   zones thought to be necessary to sustain a magnetic dynamo 
   to power a corona. Although the detection of X-rays from
   late-B and early-A type stars have been reported
   by several authors (i.e. Caillault \& Zoonematkermani 1989, 
   Schmitt et al. 1993, Bergh\"ofer \& Schmitt 1994, 
   Bergh\"ofer et al. 1996, Simon et al. 1995, Panzera et al. 1999), 
   there is no clear mechanism that explains the origin of 
   this emission. The most accepted explanation is related to the presence 
   of otherwise unknown unresolved late-type companions of these stars.

   The aim of this paper is to study the X-ray emission of 
   all L86 binary systems with PTTS candidates
   which were observed by ROSAT. For this purpose, we have selected 
   binary systems with late-type stars as secondaries. We will study 
   the X-ray emission  from the PTTS candidates  as well as the emission 
   from the early-type stars. The characteristics of the
   sample are described in Sect. 2. Sect. 3  provides 
   the details related to the source detection and identification.
   The processed X-ray data are analyzed in Sections 4, 5 and 6. 
   The  conclusions are drawn in Sect. 7.

\section{The binary sample}
 
  The Lindroos catalogue (L86) contains 78 binary systems. 
  We have selected those binaries in which the secondary member is
  a F, G, or K-type star (note that no  M-type stars are present
  in the Lindroos Catalogue). Our final sample consists of 47 systems. Two
  of them (HD\,113791 and HD\,106983) were not included in L86 but 
  in Gahm et al. (1983). We have also included the binary system
  HR\,4796 (HD\,109573, TWA 11), a possible member of the TW Hya association
  (see Webb et al. 1999) given that its stellar properties are 
  in agreement with those of the Lindroos sample. 
  This is the only binary system in our sample with an M-type star as a secondary.
  Note that we have not rejected those pairs that are classified 
  as likely optical pairs by Pallavicini et al. (1992), having in view that 
  the X-ray emission could provide important information related to the 
  nature of these systems.

   Stellar properties of our binary sample are shown in Table 1. The name of 
  the source as well as the components of the binary system 
  (A for the primary, B for the secondary and C or X for 
  companions in multiple systems) are shown in column 1. The spectral 
  types of both stars are given in column 2. 
  Column 3 shows the optical position of the primary star while column 4
  shows the projected separation between both members. 
  The distance to the primary star, the visual magnitude of both 
  components and the visual extinction to the pair are given in 
  columns 5, 6 and 7. We have also included in column 8 an orientative 
  `flag' which is related to the nature of the binary system: 
  according to the spectroscopic survey carried out by 
  Pallavicini et al. (1992),  the systems are  classified as LO 
  (likely optical pair),  PP (probably physically bound)
  and CP (certainly physically bound).

\section{ROSAT observations and data reduction}

 The X-ray telescope and the instrumentation onboard the ROSAT satellite
are described in detail by Tr\"umper (1983), Pfeffermann et al. (1988) and
David et el. (1996). Two main detectors are available: 
the Position Sensitive Proportional Counter (PSPC), which can be used
either in survey or in pointed mode, and the High Resolution Imager (HRI).
 
 We have studied all the available data for our sample, i.e., 
PSPC (survey and pointed mode) and HRI data. The source detection and 
identification have been carried out using the source detection 
routines provided by the Extended Scientific Analysis System 
(EXSAS; Zimmermann et al. 1997) which are based on a Maximum 
Likelihood (ML) technique (Cruddace et al. 1988). For all detected 
X-ray sources we have looked for optical counterparts to check
the reliability of the detection. We  will briefly describe the 
data in the following subsections.

\subsection{PSPC observations}

   \begin{table*}
   \caption{RASS Observations: Detections and Upper Limits}
    \begin{flushleft}
     \begin{tabular}{llrrrrrrr} 
      \noalign{\smallskip}
      \hline
      \noalign{\smallskip}
 
 HD & \multicolumn{2}{c}{X-ray position} & $\Delta$$^1$ & Counts$^2$ &
 Exp. time & HR1 & HR2  & ML \\
  & RA (2000) & Dec (2000) & (\arcsec)&        & (secs.) &     &      &  \\    
         \cline{2-3}
    &  (h) (m) (s) & (\degr) (\arcmin) (\arcsec) & & & & & & \\
            \noalign{\smallskip}
            \hline
            \noalign{\smallskip}

  560 & 00 10 02.73 & 11 08 49.9 & 9.3/15.9 & 115.6$\pm$11.5 & 565.4 & 
      -0.12$\pm$0.10 & 0.11$\pm$0.15 & 239.7  \\
  1438 &     &  &  & $<$3.24/3.32 & 458.3 & & & \\
  17543 & & &   &    $<$3.41/2.93 & 332.9 & & & \\ 
  23793     & 03 48 16.19 & 11 08 48.6 & 12.8/9.8 & 20.3$\pm$5.1 & 371.8 & 
   0.35$\pm$0.25 & -0.15$\pm$0.29 & 34.5 \\
  33802     & 05 12 17.81 &  -11 52 01.8 & 7.5/11.0 & 377.4$\pm$20.0 & 
  408.0 & -0.06$\pm$0.05 & -0.03$\pm$0.08 & 1124.0  \\
  35007 & & & & $<$3.26/3.36 & 431.8 & & & \\
  36151  & & & & $<$4.64/3.22 & 465.3 & & & \\
  38622     & 05 47 44.54 &  13 53 15.5 & 49.5/32.1 & 6.0 $\pm$3.6 & 
   458.5 & $>$0.21  & 0.64$\pm$0.42 & 9.4 \\
  40494      & & & & $<$6.39/5.00 & 674.9  & & &  \\
  43286   & 06 15 26.95 & 03 56 48.8 & 63.8/77.2 & 16.1$\pm$4.9 
    & 488.9 &  -0.03$\pm$0.31 & 0.55$\pm$0.34 & 18.4 \\
  48425  & & & & $<$3.49/3.41 & 522.9 & &  & \\
  53191    & & & &$<$5.74/5.14 & 1236.3 & & & \\
  56504 & &  &  &  $<$3.80/3.69& 406.4 & & & \\
  60102     & 07 11 36.51 & -84 27 00.6 & 69.2/73.4 & 11.4$\pm$4.3 & 
   345.7    & 0.76$\pm$0.35 & 0.30$\pm$0.33 & 11.4 \\
  63465   & & &  & $<$2.01/2.07  & 75.7 & & & \\
  74146   & & &  & $<$5.31/5.14& 720.1 & & & \\
  77484    & 09 02 48.18 & 00 24 17.1 & 39.0/42.1 & 7.1$\pm$3.5 & 
   372.4 & -0.97$\pm$0.39  & $<$-0.28  & 6.2 \\
  86388   &  &  &   & $<$ 3.99/4.00 & 391.3 & & & \\
  87901 & 10 08 12.72 & 12 00 01.8  & 184.8/13.5 & 48.9$\pm$7.8 & 393.6 &
   -0.37$\pm$0.15 & $<$-0.67 & 78.0\\
  90972    & 10 29 39.77 & -30 35 57.9 & 63.1/73.8 & 9.8$\pm$4.0 &
   132.8 & 0.08$\pm$0.45 & -0.24$\pm$0.50 & 6.7 \\
  104901    &  &  &  & $<$3.84/4.14 & 361.2 &  &  \\
  106983 & & &  &$<$2.16/2.16 & 82.7 & & &  \\
  108767   & 12 29 52.18 & -16 30 48.0 & 9.0/32.6 & 20.5$\pm$5.5 & 
   281.3 & -0.24$\pm$0.26 & -0.12$\pm$0.41 & 21.7 \\
  109573 & 12 35 59.5 & -39 52 05.01  & 18.9/17.2 & 43.7$\pm$7.6 & 285.0 & 
   -0.20$\pm$0.17 & 0.35$\pm$0.17 & \\
  112244    & 12 55 54.00 & -56 50 04.9 & 25.9/19.5  & 4.2$\pm$3.0 & 
   108.5 &  $>$0.46 & 0.47$\pm$0.39 & 7.0 \\
  112413    & 12 56 01.11 & 38 18 56.2  & 11.7/8.3 & 47.4$\pm$8.1 & 
   519.7 & -0.39$\pm$0.16 & 0.16$\pm$0.30 & 56.7 \\
  113703    & 13 06 18.27 & -48 27 44.3  & 16.0/3.7 & 72.2$\pm$9.1 & 
   302.7 & 0.07$\pm$0.13 & 0.00$\pm$0.17 & 150.7 \\
  113791   & 13 06 59.79  & -49 54 03.3 & 53.3/33.1 & 12.1$\pm$4.4 &
   343.3 & $>$0.45 & 0.37$\pm$0.33 & 15.7 \\  
  127304    & &  &  & $<$4.23/2.70 & 592.5 & & \\
  127971    & 14 35 31.63 & -41 31 08.0 & 5.6/21.3 & 6.3$\pm$3.7 & 
   315.0 & $>$0.26  & -0.06$\pm$0.50 & 6.8 \\
  129791   & 14 45 56.17  & -44 52 12.2 & 18.0/19.8 & 39.0$\pm$6.9 & 
   327.7 & 0.45$\pm$0.16 & -0.08$\pm$0.21 & 62.4 \\
  137387 & &  &  & $<$2.99/3.13 & 248.8  & & & \\
  138800 & &  &  & $<$2.68/2.50 & 227.8 & & & \\
  143939    & 16 04 44.67  &  -39 26 03.0 &  2.8/11.3 & 25.5$\pm$5.9 & 
   324.4 & -0.21$\pm$0.23 & -0.25$\pm$0.37 & 39.3 \\
  145483 & 16 12 16.52  &  -28 25 30.0 & 7.0/19.39  & 60.8$\pm$8.6 & 
   319.7 & 0.47$\pm$0.13 & 0.09$\pm$0.17 & 104.9 \\ 
  174585   &  &  &  & $<$2.96/3.10  & 790.1 & & & \\
  180183   &   & & & $<$ 2.47/1.98 & 139.8 & & & \\
            \noalign{\smallskip}
            \hline
   \end{tabular}
   \end{flushleft}

  {\bf Notes:}\\
  1. Given the low spatial resolution of the PSPC, we cannot resolve 
    the two components of the binary system. As a consequence, the 
    displacement $\Delta$ between the detected X-ray source and the 
    optical positions of the two
    components of the system are given. 2. In the case of non-detection, 
    upper limits are computed at the positions of both components, A and
    B, respectively. 
  
  \end{table*}

\begin{table*}
   \caption{PSPC Pointed Observations}
    \begin{flushleft}
     \begin{tabular}{l@{\hspace{2mm}}l@{\hspace{3mm}}l@{\hspace{2mm}}l
      @{\hspace{2mm}}c@{\hspace{3mm}}r@{\hspace{3mm}}r@{\hspace{3mm}}r
      @{\hspace{3mm}}r@{\hspace{3mm}}r@{\hspace{3mm}}r} 
      \noalign{\smallskip}
      \hline
      \noalign{\smallskip}
HD & ROR  & \multicolumn{2}{c}{X-ray position} & $\Delta$ & Offaxis & Counts & 
    Exp. time & HR1 & HR2  & ML \\
   & number  & RA (2000) & Dec (2000) & (\arcsec) & (\arcmin) &  & 
       (secs.) &     &      &  \\    
         \cline{3-4}
    &    & (h)  (m) (s) & (\degr)  (\arcmin)  (\arcsec) & & & & & & & \\
     \hline
     \noalign{\smallskip}
560   & 700503p & 00 10 02.48 & 11 08 41.2 & 5.6/6.8 & 12.32 & 1993.2$\pm$45.3 
      & 8238.3 & -0.04$\pm$0.02 & 0.05$\pm$0.03 & 7713.6 \\
 ''   & 701092p & 00 10 02.47 & 11 08 40.4 & 6.1/6.0 & 12.31 & 4258.3$\pm$65.7 
      & 8121.8 & -0.08$\pm$0.02 & 0.12$\pm$0.02 & 18859.8 \\
74146$^1$ & 200501p &  &  &  & 18.60 & $<$69.6/59.7 & 17319.4 &  & &   \\
104901$^2$ & 201271 & & & & 4.94 &   $<$16.2/3.94 & 8491.6 & & & \\
143939     & 200738p & 16 04 44.66 & -39 26 16.4 & 12.0/13.1 & 7.61 & 
       108.0$\pm$10.6 & 1686.6 & -0.13$\pm$0.10 & 0.09$\pm$0.14 & 334.2 \\
            \noalign{\smallskip}
            \hline
            \noalign{\smallskip}
 \end{tabular}
   \end{flushleft}

   {\bf Notes:} 1. The source is very close to the support structure 
   of the telescope, so we are not sure about the reliability 
   of the upper limit; 2. This 
   source has two PSPC observations: 201271p and 201271p-1.
   There are no detections in the individual observations, so 
   we have added up both files to improve the S/N ratio. 
   \end{table*}

The ROSAT All-Sky Survey (RASS) was performed with the PSPC.
The diameter of the field of view is 2\degr~ and each object is 
observed up to $\sim 30$ times separated by $\sim 90$
minutes, with up to $\sim$ 30 sec per scan. 

The spectral 
resolution of the PSPC (43\% 
at 0.93 keV) allows spectral analysis in three energy bands: \\

\noindent
- {\it Soft} = 0.1 to 0.4 keV \\
- {\it Hard 1}  = 0.5 to 0.9 keV \\ 
- {\it Hard 2}  = 0.9 to 2.0 keV \\

While in some pointed observations the signal to noise are 
large enough to carry out detailed spectral analysis, 
this is not possible for RASS data. Note that the 
RASS exposure times of our sample range from 75 sec to 1236 sec. 
However, we can obtain spectral information 
of our sources studying the X-ray hardness ratios (HR)
defined as follows:\\

\begin{equation} 
\; \; HR1=\frac{(H1+H2-S)}{(H1+H2+S)} \; \; \mbox{and} \;\;
      HR2=\frac{(H2-H1)}{(H2+H1)}
\end{equation}
where H1 and H2 are the counts observed in the Hard 1 
and Hard 2 bands, and S are the counts observed in the
Soft band. Hence, HR values can range from $-1$ to $+1$.
If no counts are detected in one of the bands 
only an upper or lower limit to HR1 
and HR2 is available. Neither HR values nor limits are available 
for undetected stars.

 The nominal positional accuracy of the ROSAT PSPC
detector in pointing mode is $\sim$ 25\arcsec~at 1 keV 
(note that it is energy and off-axis angle dependent). 
However, this accuracy is reduced to $\sim$ 1\arcmin~in the survey phase.
If the offset between the X-ray source and the optical position 
exceeds the spatial resolution of the RASS data ($\sim$ 1\arcmin), 
the identification of the X-ray source with the optical counterpart 
is doubtful. Hence, for all the RASS observations we have 
to select a maximal displacement between the X-ray and the optical
position  within which the detections are reliable. Following Neuh\"auser 
et al. (1995), we have taken a value of 40\arcsec~ as the maximal distance 
between the X-ray detection and the optical counterpart. 

 The RASS data (detections and upper limits) of our sample are shown in 
Table 2. The name of the source  and the 
position of the X-ray detection are given 
in columns 1 and 2.  As mentioned before, the spatial resolution of the PSPC
is too low to resolve most of the binary
systems. In fact, in most of the cases we have only obtained a 
single detection displaced from  both components
of the system. Hence, we  show in column 3 the displacements of 
the X-ray detection with respect to the optical positions of 
both members of the binary system. There is only one system in which the
projected separation is so large that the X-ray emission 
can be clearly attributed to the secondary star: HD\,87901.
The total number of counts in the broad band, the exposure time 
and the hardness ratios HR1 and HR2 are given in columns 4, 5, 6 and 7. 
In the case of non-detections we have computed the number of counts at the 
position of both components, A and B, respectively. An estimate of the
probability of the detection by the ML procedure
is given in the last column; a value of ML=5 corresponds
to a 2.7$\sigma$ signal over the background.
Note that 11 sources listed in Table 1 do not appear
in Table 2. Most  of them (HD\,8803, HD\,27638, HD\,36013, HD\,36779,
HD\,53755, HD\,71510, HD\,123445 and HD\,162082) are not detected  
and are located so close to another X-ray sources 
that a computation of an upper limit is not possible.
Two of them (HD\,70309, HD\,76566 and HD\,120641) show bad 
quality exposure maps 
so no X-ray data can be derived.

 The same information as in Table 2 is given in Table 3 for PSPC 
pointed observations. Only four of our binary systems were observed 
in PSPC pointings, with just two of them detected. 
Columns 1 and 2 provide the name of the system and the ROSAT 
Observation Request (ROR) number. The coordinates 
of the X-ray detection are given in column 3. As in the case of the 
RASS data, the spatial resolution is not high enough to detect both 
components of the binary systems. As a consequence, we show the 
displacement $\Delta$ of the X-ray source with respect to both components 
of the binary system in column 4. Column 5 shows the displacement 
of the source with respect to the center of the image (off-axis angle). 
Note that the sensitivity of the detector degrades with increasing 
off-axis angle. The total number of counts, the exposure time and 
the hardness ratios, HR1 and HR2, are given in columns 6, 7, 8 and 9. 
The ML coefficient is provided in the last column.


   \begin{table*}
   \caption{HRI Observations}
    \begin{flushleft}
     \begin{tabular}{l@{\hspace{2mm}}l@{\hspace{2mm}}c@{\hspace{4mm}}l
     @{\hspace{2mm}}r@{\hspace{3mm}}c@{\hspace{2mm}}r@{\hspace{3mm}}r
     @{\hspace{2mm}}r@{\hspace{2mm}}r@{\hspace{3mm}}r} 
      \noalign{\smallskip}
      \hline
      \noalign{\smallskip}

HD & ROR  & Comp. &\multicolumn{2}{c}{X-ray position$^1$} & $\Delta$ & 
 Offaxis & Counts & Exp. time & HR & ML\\
   & number & & RA (2000) & Dec (2000) & (\arcsec) & (\arcmin) & &(secs.) & & \\ 
     \cline{4-5}
        & &      &  (h) (m) (s) &   (\degr) (\arcmin) (\arcsec)& & & & &  & \\
            \noalign{\smallskip}
            \hline
            \noalign{\smallskip}
   
560    & 201678h  & {\bf e$^*$} A+B  & 00 10  02.39 &11 08 40.0 & 5.7/5.0 & 0.33 & 
     105.6$\pm$10.4 & 1531.7 & -0.20$\pm$0.10 & 304.3 \\ 
      & 201987h & A+B & 00 10 02.67 &  11 08 37.5 & 10.1/7.1 & 0.38 
        & 118.9$\pm$11.0 & 2373.5 & 0.58$\pm$0.07 & 505.7 \\ 
1438  & 201154h,h-1 & A+B & & & & & $<$13.51 & 5443.5 & & \\
17543 & 201331h & A & 02 49 17.63  &  17 27 47.9 & 3.8 & 0.29 &10.4$\pm$3.5 
       & 5770.1 & $<$-0.46 & 16.2 \\
      & ''     & C & 02 49 19.34 & 17 27 40.6 & 5.4 & 0.61 & 31.0$\pm$5.8 
       & 5767.9   & 0.34$\pm$0.15 & 75.1\\
27638$^s$ & 200187 & A & & & &  & $<$4.26 & 3881.6  & & \\
      & ''     & B & & &  & & $<$2.03 & 2312.9 & & \\
33802$^{2,s}$  & 200185h & {\bf e} A  & 05 12 17.95 & -11 52 05.4 & 3.8 & 
 0.30 & 608.7$\pm$22.9 & 2349.8 & & 2500.5 \\
       & ''     & {\bf e} B &05 12 17.55& -11 51 56.5 & 4.5 & 0.32 & 77.8$\pm$8.8 & 
      2350.1 &       & 352.6 \\  
38622  & 201333h & A & 05 47 42.63 &  13 53 56.2 & 4.5 & 0.16 & 15.1$\pm$4.0 
        & 4139.7 & $<$-0.43 & 33.8 \\
       & ''     & C & 05 47 42.55 & 13 53 31.9 & 2.1 & 0.50 & 10.9$\pm$3.5
        & 4139.0 & $>$0.54 &  21.3 \\
40494  & 201685h & A & 05 57 32.14 &  -35 16 59.6 & 0.8 & 0.28 & 32.7$\pm$5.9 
       & 4539.2 & $<$-0.78 & 89.4 \\
       & '' & B  & & & &  & $<$1.59 & 4535.3.8 & & \\
53191$^3$ & 201684h & B & 07 00 14.53 &  -60 51 54.4 & 0.7  & 0.41 & 
  5.9 $\pm$2.6  & 4813.5 &  & 7.9\\ 
60102$^3$ & 201328h & B & 07 11 27.90 & -84 28 10.3 
       & 2.6 & 0.25 & 28.6$\pm$5.4 & 2316.0 & 0.45$\pm$0.17 & 101.8\\
74146$^4$   & 201447h & A & 08 39 57.41 & -53 03 15.3 & 2.5 & 16.02& 
       172.4$\pm$19.0 & 50034.5& $<$-0.68 & 69.9\\
      & 201682h & A & 08 39 57.78  &  -53 03 15.2 & 2.5 & 0.25 & 10.3$\pm$3.3 
        & 4680.5& -0.40$\pm$0.11 & 24.1 \\
86388$^3$ & 201683h & A+B?,B  & 09 55 04.26 & -69 11 28.0 & 11.4/6.4 & 0.08 & 
        31.3$\pm$6.1 & 2200.6 & -0.04$\pm$0.19 & 51.4 \\
87901  & 800807h & A  & 10 08 22.37 & 11 58 05.8 & 4.0 & 11.59 & 173.4$\pm$13.6 
       &2781.5 &-0.34 $\pm$0.07  & 460.0\\
       & `` & B & 10 08 12.73 & 11 59 50.7 & 2.8 & 14.32 & 43.2$\pm$7.6 & 
       2712.7 & 0.42$\pm$0.16 & 44.6\\
       & 800807h-1& A & 10 08 22.53 & 11 58 02.2 & 3.3 & 11.52 & 
        1221.5$\pm$35.5 & 18262.3 & -0.03$\pm$0.03 & 3795.3\\
       & ''       & B &10 08 12.95 & 11 59 47.8 & 1.9 & 14.25& 357.8$\pm$20.8 
        & 17774.6 & 0.69$\pm$0.04 & 506.1 \\
       & 800880h & A & 10 08 22.10 & 11 58 08.5 & 7.2 & 12.47 
         & 4565.8$\pm$69.0 & 60144.2 & -0.36$\pm$0.01 &  9914.8 \\
       & '' &  B & 10 08 12.64& 11 59 55.7 & 7.8 & 15.23 & 
           1191.1$\pm$38.4 & 5853.6 & 0.41$\pm$0.03 & 1560.7 \\
90972$^{3,s}$ & 200188h & B & 10 29 34.59 & -30 36 30.4 & 2.0 & 0.32 & 
  44.4$\pm$6.9 & 4524.8 & 0.21$\pm$0.15 & 120.3 \\ 
108767 & 201329h &  A & 12 29 52.11 & -16 30 54.3 & 3.9 & 0.34 & 40.3$\pm$6.5 
        & 3772.6 & -0.47$\pm$0.14 & 114.1\\
       & ''     &  B  & 12 29 51.07 & -16 31 14.2 & 2.5 & 0.42 & 102.3$\pm$10.2
        & 3775.6 & 0.60$\pm$0.08 & 384.0\\
       & 201679h &  A & 12 29 52.03 & -16 30 54.7 & 2.7 & 0.35 & 43.7$\pm$6.7 
        & 4277.6 & -0.21$\pm$0.15 & 135.2\\
       & ''     &  B  & 12 29 50.97 & -16 31 14.4 & 1.1 & 0.45 & 91.7$\pm$9.8 
        & 4279.1 & 0.62$\pm$0.08 & 278.4 \\
       & 201991h & A & 12 29 51.70 &  -16 30 53.3 & 3.1 & 0.39 & 24.8$\pm$5.1 
        & 2432.4 & -0.34$\pm$0.20 & 60.8 \\
       & ''     & B  &12 29 50.67  & -16 31 13.6 & 3.6 & 0.81 & 40.5$\pm$6.5 
        & 2431.8 & 0.55$\pm$ 0.13 & 104.3 \\
       & 201991h-1 & A & 12 29 51.95 &  -16 30 53.7 & 2.3&0.36 & 40.8$\pm$6.7 
       & 4218.0 & $<$0.01 & 68.5 \\
       & ''  & B & 12 29 51.10 & -16 31 13.4 & 3.2 & 0.75 & 113.6$\pm$10.8 
       & 4218.3 & 0.52$\pm$0.08 & 336.3 \\
109573 & 702764h & {\bf e} A+B,B? & 12 35 56.35  & -39 50 15.5 & 7.1/4.7 & 
     2.69 & 942.4$\pm$31.1 &
       40293.9 & 0.55$\pm$0.03 &  3439.0 \\
113703$^{3,s}$ & 150034h & {\bf e} A+B,B? & 13 06 18.39 & -48 27 45.2 & 17.0/5.0 &
        0.30 &  106.5$\pm$9.1 & 3042.1 & 0.01$\pm$0.09 & 412.4\\
113791 & 201680h & A & 13 06 54.9 & -49 54 22.7 & 2.6 & 0.23 & 107.5$\pm$10.5 
       & 6102.4 & -0.07$\pm$0.10 & 358.2 \\
       & '' &   B & 13 06 57.37  & -49 54 27.3 & 2.9 & 0.38 & 68.5$\pm$8.5
       & 6101.9 & 0.69$\pm$0.09 & 173.6 \\
123445 & 200183h & A & 14 08 52.24& -43 28 09.5 & 6.6 & 0.11 &50.36$\pm$7.2 
       & 4100.6 & 0.19$\pm$0.14 & 161.1 \\
      & ''     & B & & & & & $<2.0$ & 4099.0 & &  \\
127304 & 201681h & A   & & & & & $<$4.48 &5158.0 & & \\
       & 201681h & B & 14 29 48.08 & 31 47 22.3 & 5.4 & 0.50 & 5.7$\pm$2.7 
       & 5158.4 & & 7.7 \\
127971$^s$ & 200182h & A & 14 35 31.30 & -41 31 05.5 & 3.4 & 0.11  & 44.5$\pm$6.8 
       &5420.1 & 0.49$\pm$0.13 & 146.8\\
       & ''     & B & & & & & $<$1.46 & 12012.1 & & \\
129791$^s$ & 200184h & A & 14 45 57.61 & -44 52 05.6& 2.5 & 0.11 & 163.1$\pm$12.9 
       & 5063.0 & 0.56$\pm$0.07 & 684.6 \\
       & ''     & B & 14 45 56.21 & -44 52 36.8 & 4.9 & 0.48 & 24.7$\pm$5.1
       & 5061.4 & $>$0.77& 60.5 \\
143939$^{3,s}$ & 200190h & B & 16 04 43.98& -39 26 14.9& 5.1& 0.21 &
      168.6$\pm$13.1 & 7841.5 & 0.45$\pm$0.07 & 609.0 \\ 
   \noalign{\smallskip} \hline
   \end{tabular}
   \end{flushleft}

  {\bf Notes:} 1. If both components are detected, coordinates 
   of both X-ray sources are given. 
   The displacement of the secondary with respect to the
   detected X-ray source is computed comparing with its optical 
   coordinates;
2. The two stars of the system are not completely resolved so each star 
   is contributing to the count rate of the companion. 
3. The undetected component is so close to the X-ray source 
   that no reliable upper limit can be computed;
4. HD74146 has another observation, 200186h, but with very low 
   exposure time (t$_{exp}$ = 357.5 s);
{\bf e} for elongated sources; $^*$The elongated shape is due to attitude problems in the processed data;
$^s$ stars previously studied by Schmitt et al. (1993).

\end{table*}

\subsection{HRI observations}

\begin{figure*}
\vspace{0.5cm}
\begin{center}
\resizebox{15cm}{!}{\includegraphics{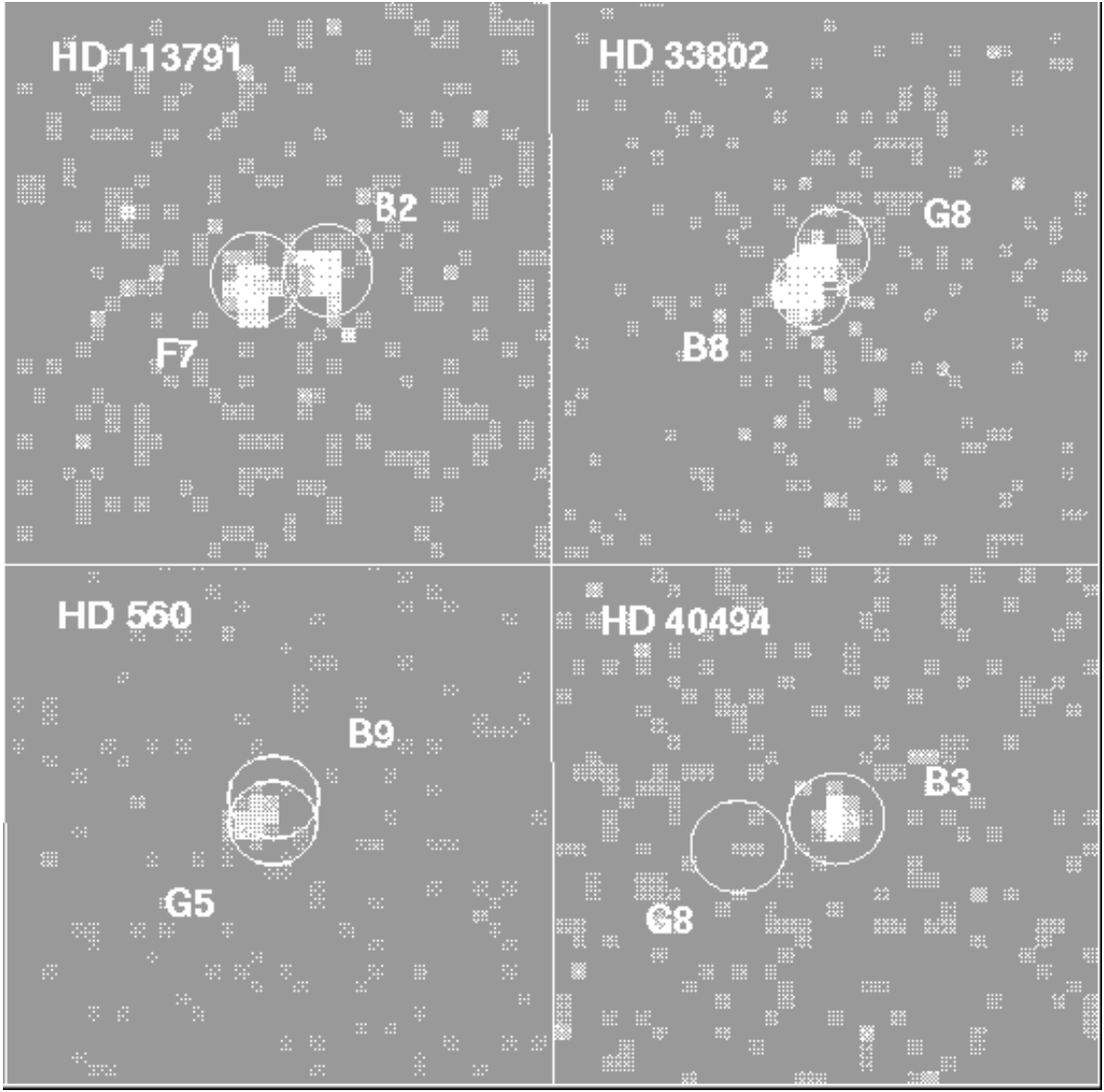}}
      \vspace{0.5cm}
      \caption{ROSAT HRI images of four Lindroos binary systems: HD\,113791, 
       HD\,33802, HD\,560 and HD\,40494. We show different cases of detection:
       HD\,113791: both components are detected and resolved, HD\,33802: both 
       components are detected but not completely resolved, HD\,560: 
       the binary is detected but unresolved and HD\,40494: 
       the binary is resolvable but only one of the members is detected.}
\end{center}
\end{figure*}

 The HRI detector allows high resolution imaging of  X-rays sources. 
The nominal spatial resolution of the detector is 1.7\arcsec~but a bore-sight 
correction as much as 10\arcsec~should be applied. The projected separations 
between the members of our binary systems range from 4.4\arcsec~to 177\arcsec. 
Most of them show separations larger than 10\arcsec~ so, in principle, 
they can be resolved by the HRI. 

The spectral resolution of the HRI is worse than that of the PSPC. However, 
the HRI pulse height distribution can be used to compute a 2-band (H and S) 
hardness ratio. We  have defined the HRI hardness ratio as:
\begin{equation}
HR = \frac{H-S}{H+S}
\end{equation}

To obtain the most realiable S- and H-band definitions, 
several ROSAT sources known from their PSPC pointed observations 
to be either extremely soft (HR1 $<$ -0.8)
or extremely hard (HR1 $>$ +0.8) were 
selected ($\sim$20). For these sources the  HRI hardness ratios
were calculated according to Eq. (2) for the following different 
soft and hard band definitions: S1 = 0-5, H1 = 6-15; S2 = 0-4, H2 = 5-15;
S3 = 0-3, H3 = 4-15; and S4 = 0-2, H4 = 3-15, where the numbers give the
corresponding HRI energy channels. For each of the corresponding four 
HRI hardness ratios the mean difference to the PSPC hardness ratios (HR1), 
was computed. The HRI bands for which the selected sources show 
the smallest deviation to the PSPC hardness ratio were S3,H3. 
Therefore,  the HRI bands are defined as the soft band (S-band) 
corresponding to channels 0 to 3 
(E $\la$ 0.3 keV) and the hard band (H-band) corresponding 
to channels 4 to 15 (E $\ga$ 0.3 keV).
This HRI hardness ratio shows a smaller dynamical range than 
the PSPC hardness ratio, HR1, but clearly identifies soft sources 
with negative values and hard sources with positive
values. For more details, see Supper et al. (in prep.).

 Table 4 shows the HRI detections of Lindroos binaries. Most
of the HRI pointed observations were carried out by one of us (HZ) as 
Principle Investigator (PI). Columns 1 to 3 give the name of 
the source, the ROR number, and the component of the system detected: 
A+B when both stars are not resolved and there is only a single detection, 
and A or B (or C or X for secondaries in multiple systems) when the members of 
the binary system have been resolved. We have also added
an 'e' to designate the elongated sources. The position of the X-ray
sources are shown in column 4. Column 5 shows the difference between 
the optical and X-ray position. Note that the optical coordinates 
used as a reference for the B-components are those  of the secondaries 
(taken from SIMBAD database). Column 6 shows the displacement of the 
sources with respect to the axis of the telescope. Finally, the total 
counts in the broad band, the exposure time, the hardness
ratio and the ML coefficient are given in columns 7, 8, 9 and 10.
The upper limits of the undetected sources are also shown.

Table 4 includes three binary systems (HD\,86388, HD\,109573 
and HD\,113703) for which it is not clear which of the two members 
of the system is responsible for the X-ray emission. 
After the comparison of the optical and the X-ray position,
we believe that the late-type companion is most likely the X-ray emitter
in the three cases. This is clearer for the HD\,86388 system 
because the X-ray detection is not elongated and it is closer 
to the optical position of the secondary. 
In the case of HD\,109573 and HD\,113703 the 
X-ray detections in the broad band are slightly 
elongated although much closer to the optical 
position of the secondary star (see Table 4).
Hence, we will assume that the X-ray emission comes from the late-type
secondaries in the three cases under study.
  
In order to illustrate different observations of the 
Lindroos binary systems, we  have shown in Fig. 1 several HRI images of 
different pairs. 


\begin{table}
 \caption{X-ray luminosities}
 \begin{flushleft}
  \begin{tabular}{l@{\hspace{2mm}}ll@{\hspace{3mm}}ll} 
            \hline

HD & Comp. & $\lg{N_{\rm H}}$  & Det.& $\lg{L_{\rm x}}$ \\
   &  & (cm$^{-2}$)   &  &   (erg/s) \\ 
            \hline
            \noalign{\smallskip}
560    & A+B &  19.74 & H   & 30.56$\pm$29.41$^*$  \\    
1438   & A+B & '' & H  & $<$29.79   \\ 
17543  & A   & 20.57 & H  & 29.64$\pm$29.16 \\
17543  & B   & '' &  H & 30.17$\pm$29.44\\
23793  & A+B & 20.15 & R  & 30.25$\pm$29.65  \\
27638  & A   & 18.00 & H  & $<$28.56   \\
27638  & B   & ''    & H  & $<$28.49   \\ 
33802  & A   & 19.25 & H    & 30.89$\pm$29.47 \\
33802  & B   & ''    & H    & 30.03$\pm$29.06 \\
35007  & A,B & 20.46  & R & $<$29.5,$<$29.98  \\
36151  & A,B & 20.16  & R & $<$30.14,$<$30.00  \\
38622  & A   & 19.85  & H & 30.08$\pm$29.48 \\
38622  & B   &   ''   & H & 29.99$\pm$29.48 \\
40494  & A   & 18.00  & H  & 30.39$\pm$29.64 \\
40494  & B   & ''  & H  & $<$28.57  \\
43286  & A+B & 20.03  & R  & 30.52$\pm$30.01 \\
48425  & A,B & 18.00  & R  &  $<$29.94,$<$29.98  \\
53191  & B   & 19.95  & H & 29.53$\pm$29.15  \\
56504  & A,B &  20.09 & R & $<$30.92,30.93   \\
60102  & B   & 20.59  & H  & 30.62$\pm$29.89 \\
63465  & A,B & 20.65& R &  $<$30.61,$<$30.65   \\
74146  & A   & 19.73& H & 29.42$\pm$28.57$^*$    \\
77484  & A+B & 20.20 & R  & 30.11$\pm$29.81 \\
86388  & B:  & 19.55  & H & 29.42$\pm$28.75 \\
87901  & A & $<$18.08$^2$ & H & 27.73$\pm$26.15$^*$  \\
87901  & B & '' & H   & 28.76$\pm$27.50$^*$  \\
90972  & B   & 19.73 & H & 30.11$\pm$29.29 \\
104901 & A+B & 21.29 & P & $<$30.32  \\
106953 & A+B & 18.00 & R & $<$29.48,$<$29.53\\
108767 & A & 18.38$^3$& H  & 28.58$\pm$27.45$^*$   \\
108767 & B & '' & H  & 28.95$\pm$27.69$^*$   \\
109573 & B & 18.00 &  H & 29.75$\pm$28.26 \\
112244 &A+B & 21.20 & R & 31.18$\pm$31.01 \\
112413 &A+B & $<$18.11$^2$ & R & 29.04$\pm$28.27 \\
113703 & B   & 20.24 & H &  30.57$\pm$29.51 \\
113791 & A   & 18.00 & H &  30.14$\pm$29.13  \\
113791 & B   & ''   & H &  29.98$\pm$29.07  \\
123445 & A   & 20.33 & H &  30.54$\pm$29.71  \\
123445 & B   & ''    & H &  $<$29.21  \\
127304 & A   & 19.85 & H &  $<$28.73  \\
127304 & B   &  ''   & H &  28.89$\pm$28.55 \\
127971 & A   & 20.15 & H &  29.74$\pm$28.91 \\
127971 & B   &   ''  & H &  $<$27.90   \\
129791 & A   & 20.66 & H &  30.58$\pm$29.47 \\
129791 & B   &  ''   & H &  30.63$\pm$29.52 \\
137387 & A,B & 20.92 & R  & $<$30.10,$<$30.16  \\
138800 & A,B & 20.59 & R  & $<$29.81,$<$29.80 \\
143939 & B   & 18.00 & H & 30.50$\pm$29.40 \\
145483 & A+B & 20.65 & R & 30.26$\pm$29.41\\
174585 & A+B & 20.48 & R & $<$29.59,$<$29.63  \\
180183 & A+B & 20.25 & R & $<$30.03,$<$29.96  \\
\hline
\end{tabular}
\end{flushleft}
{\bf Notes:} 1. Adopted from Bergh\"ofer et al. (1996);
2. Adopted from Fruscione et al. (1994); 
$^*$Mean X-ray luminosities computed from several observations. 
\end{table}


\subsection{X-ray fluxes and luminosities}

 The X-rays fluxes can be computed by multiplying the observed count
rates by an energy conversion factor (ECF).  The ECF depends on the
detector response and the underlying model for the X-ray spectrum.  In
the case of our sample we have assumed a 1-T Raymond-Smith thermal
spectrum (Raymond \& Smith 1977), which implies that the ECF mainly
depends on the temperature of the emitting plasma and the interstellar
absorption.  The assumed temperature for the late-type stars is $ kT_x
= 1 \; {\rm keV} $ (with $ k = $ Boltzmann's constant) which is suitable for
active late-type stars such as TTS (see Neuh\"auser et al. 1995). 
In the case of the B-type stars we have adopted a mean value of 
$kT_x = 0.5 \; {\rm keV}$ (see Bergh\"ofer \& Schmitt 1994 
and Bergh\"ofer et al. 1996).

 In order to correct for the interstellar absorption, we have converted
the visual extinction $A_v$ to our sources into hydrogen column
densities, $N_H$, following Paresce (1984):

\begin{equation}
\frac{N_H}{{\rm cm^2}}= 5.5 \cdot 10^{21}\frac{E(B-V)}{mag} =
   \frac{5.5}{3.1} \cdot 10^{21} \, \frac{A_v}{mag}
\end{equation}

For those cases for which no $A_v$ is available,
we have adopted a lower limit of log $(N_H/cm^{-2})=18$. 

The PSPC ECF for different temperatures and
interstellar absorption column densities are provided by 
Neuh\"auser et al. (1995) for late-type stars, and by Bergh\"ofer et
al. (1996) for early-type stars. In the case of the HRI 
observations, we have computed the ECF's following the Technical 
Appendix to the {\it ROSAT Call for proposals}.

Once we obtain the X-ray fluxes,  the X-ray luminosities are given by
\begin{equation}
 L_x = 4 \cdot \pi \cdot d^2 \cdot f_x 
\end{equation} 
with  $d$ being the distance to the star.  We have made use of the
Hipparcos parallaxes of the primary stars to estimate the distances to
our sources (see Table 1).
\begin{figure*}[t]
\resizebox{18cm}{!}{\includegraphics{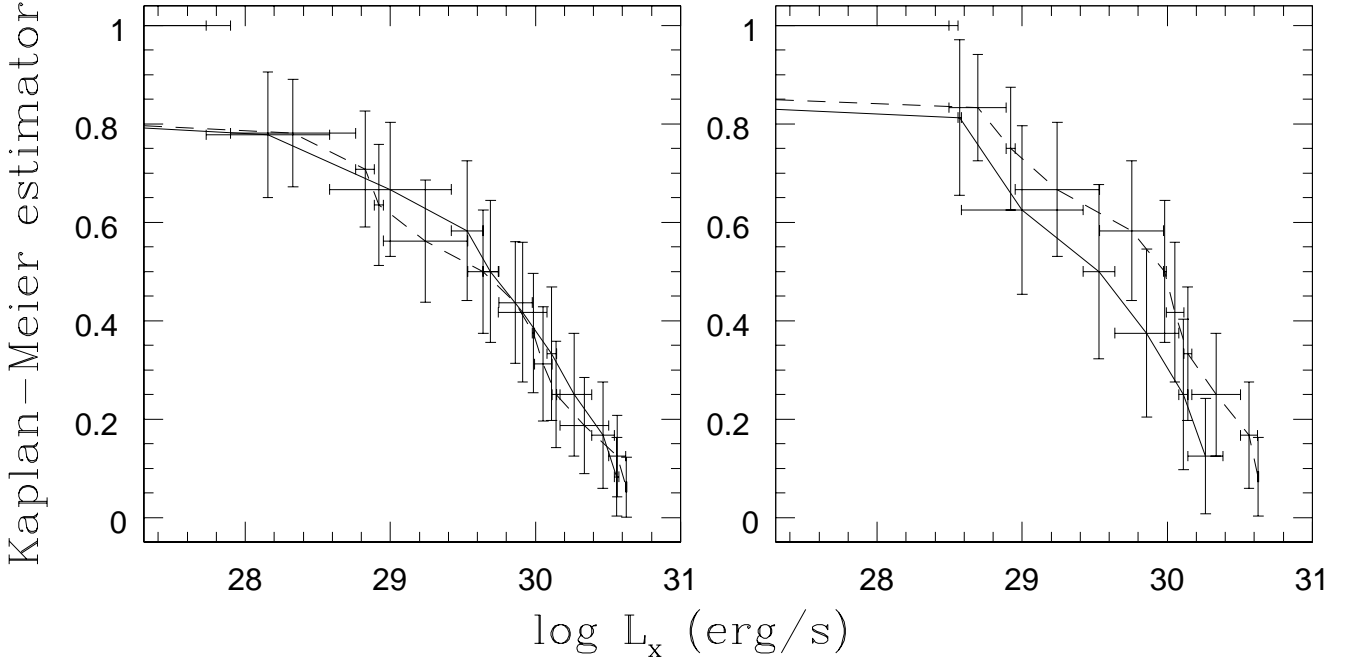}}
\vspace{-9.0cm}
\caption{X-ray luminosity functions (XLF): Kaplan-Meier estimator
versus X-ray luminosity for both early- and late-type stars.
The left panel contains a sample of stars in 
resolved systems while the right panel does not contain stars
that could be contaminating the samples for different reasons (see text).
The early- and late-type 
star data are connected by a solid and a dashed line respectively.
We can see that the XLF is the same for both cases.
Although the resulting probabilities vary from one plot to the other,
the main result is that our samples 
are statistically similar in the two cases.}
\end{figure*}

 The X-ray fluxes and luminosities for the whole sample are given 
in Table 5. The HD number and the binary component (A, B, C or X) are shown 
in columns 1 and 2.  Column  3 provides the computed hydrogen column density.  
The type of observation, R for RASS, P for PSPC and H for HRI, 
is given in column 4. The X-ray luminosities with 
their respective errors and the upper limits for non-detections
are finally listed in the last column.
If the binary system is observed and resolved  
by the HRI, we provide the fluxes and luminosities for both 
members of the pair. If unresolved, we just can provide
one X-ray luminosity associated with the pair. 
In these cases we have adopted a temperature of 1 $keV$ 
to compute the ECF's although, in principle, 
both stars could be contributing to the total emission.
The last method has also been applied to those systems with
only RASS or PSPC pointed observations, given that 
they are always  unresolved. In the
case of RASS or PSPC non-detections, we have estimated two
upper limit luminosities, corresponding to the count rates computed 
at the optical positions of the primary and
the secondary star, respectively (see Table 2).  

\section{Interpretation of the data}

The PSPC is not able to resolve most of the binary systems
of the sample (the only exception is HD\,87901). 
Therefore, the  HRI data are the most appropriate 
to study the X-ray emission of individual stars
in Lindroos systems. We have based 
the following analysis on the HRI data of resolved pairs. 
As we will show in Section 5, most of the resolved pairs 
show evidences of being physically bound. 
Only two binary systems are definitely optical pairs 
(HD\,123445 and HD\,127971), while another two systems are doubtful
(HD\,40494 and HD\,87901).

\subsection{X-ray luminosity functions}

The X-ray luminosity function (XLF)  of a sample can be derived with 
Kaplan-Meier estimators using the statistical package ASURV 
(see Feigelson \& Nelson 1985, Schmitt, 1985 and Isobe et al. 1986), 
which allows to take into account both detections and upper limits. 

 To check whether the two samples of early-type primaries and
late-type secondaries are statistically different
we have performed a two-sample test with ASURV.  
We do not include those objects which are either unresolved 
or whose identification is not clear, namely: HD\,560 and HD\,1438, 
HD\,33802, HD\,109573, HD\,113703 and HD\,86388. 
The result of the test is that our samples are statistically 
similar with a probability
of 0.8 (see upper panel of Fig. 2). Given the different spectral types 
of the stars 
in the two samples, this conclusion may appear surprising. 
In order to check  this result, we have repeated the statistical 
analysis after removing those sources that could be 
contaminating both samples. In the case of the late-type stars 
we have removed the X-ray upper limits of two probably 
unbound sources, HD\,123445\,B, HD\,127971\,B. 
In the case of early-type stars we have removed 
one source with an unreliable X-ray detection ( HD\,87901\,A,  see
Sect. 4.2), and three sources for which we found indications of  
unresolved late-type companions (Sect. 4.3): HD\,123445\,A, HD\,127971\,A 
and HD\,129791\,A.  
For this reduced sample, the probability of both groups
of stars to be statistically similar is reduced to 0.4 but this value
is higher than the threshold (0.05) to reject 
the null hypothesis of our two samples to be equivalent.
(see lower panel of Fig. 2).

Note, however, that the X-ray luminosity alone cannot provide information 
about the nature of the X-ray emission. This could only be studied 
through the spectral analysis of the emission or through the 
hardness ratios (HR's).

\subsection{Hardness ratios}

In the case of the HRI observations, we could obtain
individual hardness ratios (HR) using the two sets of channels defined
in Sect. 3.2.  As shown in Table 4, the computed hardness ratios 
generally differ from early- to late-type stars.  In most of the
cases, late-type stars show positive HR's while early-type stars
show negative values. This means that late-type stars emit most of
their X-ray radiation in the H-band, which is consistent with the
presence of an energetic corona, while early-type stars mainly emit in
the S-band. In fact, there are several stars which are only detected
in one of these bands, showing upper or lower limits to the HR's.
However, there are some stars which do not follow this trend: 
HD\,87901A, HD\,123445A, HD\,127971A and HD\,129791A,
all of which are early-type stars so they are not supposed
to have an energetic corona which could explain their positive HR's.

\begin{table}
\begin{flushleft}
\caption{HRI Soft-band rate and UV-rate for detected
  early-type stars.}
\begin{tabular}{l@{\hspace{1.5mm}}cr@{\hspace{1.5mm}}rr@{\hspace{1.5mm}}c}
\hline
 HD & U & S-rate  & UV-rate & 
  (S/UV) & $log(\frac{L_x}{L_{bol}}$)   \\ 
       & (mag) & (cts/ks)      & (cts/ks) &       &  \\ \hline
17543  & 4.69  & 1.81          & 0.24     & 7.54  &   -6.8 \\
38622  & 4.46  & 3.71          & 0.32     &  4.46 &   -7.0 \\
40494  & 3.52  & 7.26          & 1.06     & 6.48  &   -6.9 \\
74146  & 4.48  & 2.86          & 0.32     & 8.93  &   -6.8 \\
87901  & 0.88  & 68.35         & 30.87    & 2.21  &   -8.3 \\
108767 & 2.81  & 7.89          & 2.62     & 3.01  &   -6.8 \\ 
113791 & 3.28  & 17.59         & 1.43     & 12.30 &   -6.6 \\ 
123445$^*$ & 6.02 & 12.4  &  0.09 & 286.11        &   -5.4 \\
127971$^*$ & 5.39 & 7.90  &  0.04 & 81.48         &   -5.9 \\
129791$^*$ & 6.84 & 31.44 &  0.01 & 2069.78   &  -4.6 \\ \hline
\end{tabular}
\end{flushleft}
{\bf Notes:} 
$^*$ No U-magnitudes available. We
derived them from their spectral types and their V-magnitudes, 
according to Kenyon \& Hartmann (1995). 
\end{table}

Bergh\"ofer et al. (1999) has shown that the HRI detector is sensitive
to ultraviolet (UV) radiation below 4000\AA. The contamination of the
UV light to the final count rate is mainly concentrated to pulse
height channels 1-3 ($\sim$ our S-band range). Our primary stars are
bright sources in the UV range so, in principle, the computed S-band
rate could be just a response of the detector to the UV light.  In
order to check the reliability of the S-band counts for the early-type
stars, we have estimated the contribution of the photospheric UV light
to the total S-band rate. Following Bergh\"ofer et al. (1999):
\begin{equation} 
 HRI_{UV} = 10^{(-1.022\pm0.003)-(0.555\pm0.005)U} \; \; \;(cts/s) 
\end{equation}
with U being the U-magnitudes of the observed stars.  This equation was
deduced considering the emission of the sources from channels 1 to 8
of the HRI detector.  In order to be conservative, we have also
computed the S-band rates in these channels. Table 6 shows the
results.  Column 1 provides the name of the star. The U-magnitudes,
taken from {\it The Bright Star Catalogue} (Hoffleit \& Jaschek 1991)
are shown in column 2. For three sources U-magnitudes are not 
available, so we deduced them from their spectral types
and V-magnitudes following Kenyon \& Hartmann (1995).
The S-band rate and the UV rate, both computed
in channels 1-8 of the detector, are shown in columns 3 and 4, while
their ratio is provided column 5. We finally show the 
derived $\lg({L_{\rm x}}/{L_{\rm bol}})$ ratio in the 
last column, in order to
check the reliability of the X-ray detections. The 
bolometric luminosities have been computed using the stellar data
provided in Table 1 and the bolometric corrections from Schmidt-Kaler (1982).

As we can see from Table 6, most of the B-type stars of our sample
have S-band rates significantly higher than the estimated UV-rates. 
There is only one source for which both values are comparable: HD\,87901A. 
Note that the HRI observations of this star show very large 
off-axis values and, according to Bergh\"ofer et al. (1999), Eq. (1) 
is not reliable for these cases. However, this system 
was resolved by the PSPC in the RASS survey (see Tab. 2) and, as we mentioned 
in Sect. 3, no emission was detected from the early-type star but 
from the secondary. 

To further test the reliability of the
S-band emission in early-type stars, we have also studied the 
$\lg({L_{\rm x}}/{L_{\rm bol}})$ ratio of these stars to see 
if it is consistent with the
ratio for stars of similar spectral types. 
According to Table 6, most of the stars
show ratios consistent with the reported values for 
early-type stars (Bergh\"ofer et al. 1997). 
There are also four stars with unusual ratios: HD\,87901
with a lower ratio, and HD\,123445, HD\,127971 and HD\,129791, with 
ratios closer to those found in late-type stars.
Given that these three early-type stars also show positive HR, 
it is possible that they have unresolved late-type companions. 
However, in the case of HD\,87901 the HR is negative and the 
$\lg({L_{\rm x}}/{L_{\rm bol}})$ is lower than in late-type stars, 
so the X-ray emission can not be related with an unresolved 
source. This fact together with its non-detection in the RASS
make us think that this emission is due to the HRI UV leak.
Therefore, excluding HD\,87901\,A, we can identify 
the computed S-band rates with intrinsic X-ray
emission from the sources. 

The previous test allows to confirm the reliability of the computed
HRI HR's. Given that these HR's are systematically positive for
late-type stars and systematically negative for early-type stars, we
can conclude that the nature of the X-ray emission is intrinsically
different for our two samples.  We will discuss this point more deeply
in the following subsection.
   \begin{figure}
     \vspace{-0.5cm}
     \resizebox{9cm}{!}{\includegraphics{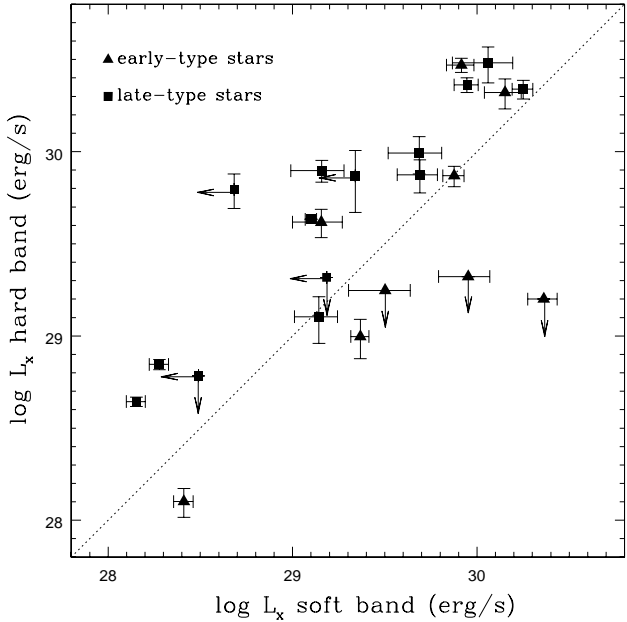}}
      \caption{X-ray luminosities in the soft and hard bands for the HRI 
           X-ray detections. As a reference, we have overplotted
           a 1:1 correlation (dotted line). The early-type primaries 
           are represented by filled triangles while the late-type 
           secondaries are represented 
           by filled squares. Late-type stars show
           generally harder X-ray emission
           than early-type stars.}
   \end{figure}

\subsection{X-ray luminosities}

The analysis of the HR's suggests that the nature of the X-ray 
emission is different for our two samples. A useful way to represent 
this difference 
consists of comparing the X-ray luminosities obtained in the two HRI bands, 
S and H, for all of our detected sources.  Fig. 3 shows this comparison. Note
that we have converted the S- and H-band rates into luminosities
following the procedure described in Sect. 3.3.  
We have not included HD\,33802 because the binary is not completely resolved 
and it is difficult to obtain reliable measurements from
the individual members (see Fig. 1).
 \begin{figure*}
     \vspace{-1.2cm}
     \resizebox{18cm}{!}{\includegraphics{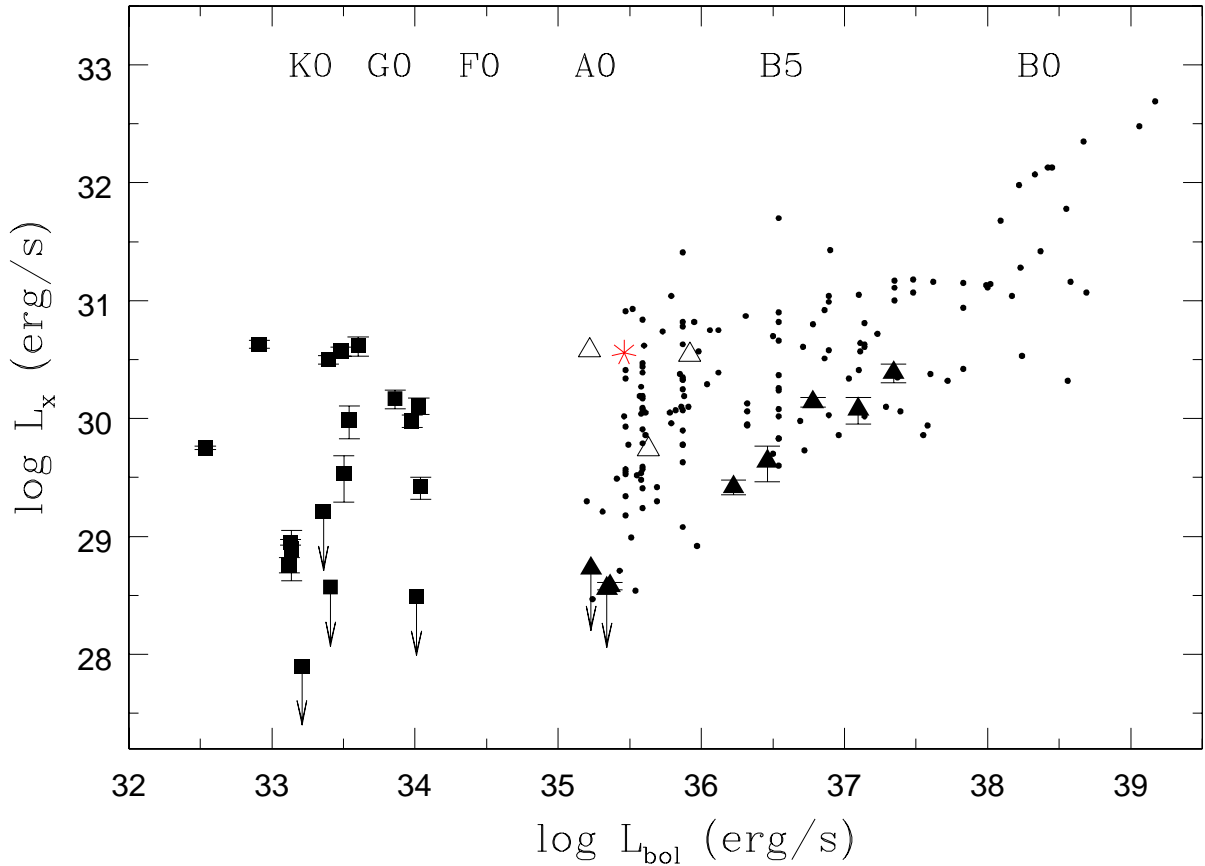}}
     \vspace{-5cm}
     \caption{Bolometric luminosity vs. X-ray  luminosity of the 
        individual stars of our sample. 
        The filled triangles correspond to the early-type primaries and the 
        filled squares to the late-type companions. The three open triangles 
        in the middle of the figure represent the Lindroos primaries with 
        possible unresolved late-type companions while the starred 
        symbol corresponds to HD\,560 (see text). We have also overplotted 
        a sample of main sequence B-type 
        stars extracted from Bergh\"ofer et al. (1996). They are represented
        by filled circles.} 
                \label{Lbol}
   \end{figure*}

As can be seen from Fig. 3, there is a clear separation between
early- and late-type stars. Early-type stars are generally softer
than the late-type companions. This result is in agreement with that obtained 
from the HR's analysis. Note that although most of the early-type stars 
lie below the 1:1 correlation line, there are three early-type primaries 
with higher X-ray luminosities in the H-band than in the S-band:
HD\,123445\,A, HD\,127971\,A and HD\,129791\,A.  As we explained
before, the higher H-band luminosities obtained for the later-type
stars are in agreement with their condition of active stars (note
that all are located above or on the 1:1 correlation line). In
contrast, the higher H-band luminosities for the early type stars are
not easy to explain, given that these stars are not supposed to
have a convective zone able to support a corona. Because these
stars lie in the same part of the diagram as the late-type secondaries, 
the simplest explanation is to relate them to unresolved late-type 
companions.  
  
   \begin{figure*}
   \vspace{-0.8cm}  
     \resizebox{12cm}{!}{\includegraphics{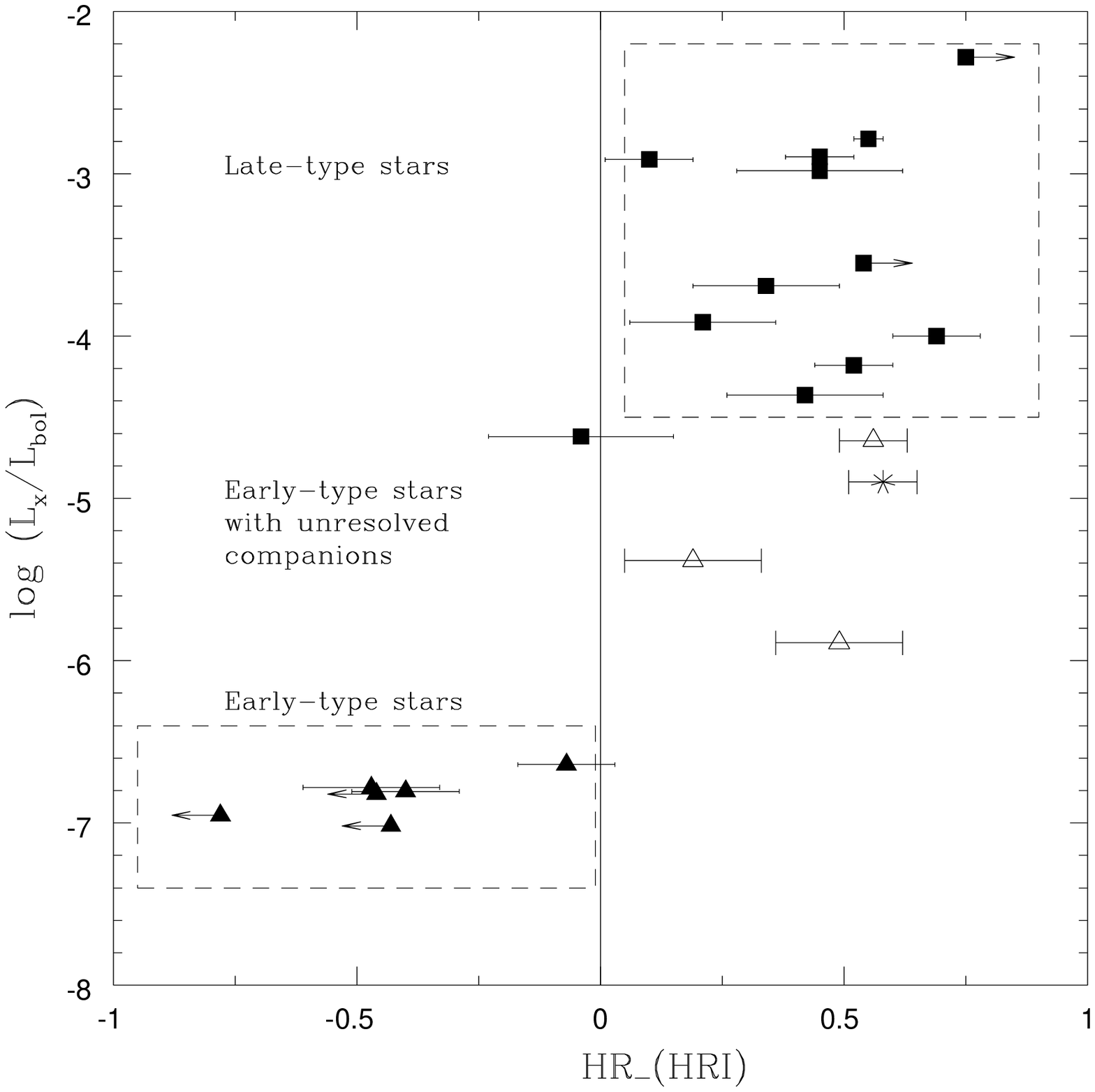}}
     \hfill
     \parbox[b]{55mm}{
     \caption{HRI Hardness ratios vs. the ratio of the
        X-ray to the bolometric luminosity. The symbols
        are the same as in Fig. 4. A solid line
        divides the soft and hard emitters. We have also plotted 
        boxes containing two different groups. The late-type stars
        lie at the right top corner of the figure while the  
        early-type stars lie in the opposite corner.
        It seems that the stars with unresolved late-type candidates
        lie in the middle of the plot with positive HR's. Also  
        HD\,560 is represented by a starred symbol (see text) lying in the same
        region as HD\,123445\,A, HD\,127971\,A and HD\,129791\,A.}
     \label{Lbol}}
   \end{figure*}
   
After the analysis of the X-ray luminosities in the two bands,
we have studied the total X-ray luminosities of
the whole sample. Given that the X-ray emission typically assumes
a particular value for each spectral type, we have plotted the X-ray
luminosity against the bolometric luminosity for each star. 
The bolometric corrections were adopted from Schmidt-Kaler (1982).
Only in the case of the M-type star, HD\,109573\,B 
(HR4796\,B), the bolometric correction was taken from 
Kenyon \& Hartmann (1995).

   As shown in Fig. 4, there is a  clear separation between late- and 
early-type stars. Late-type stars lie in the region of
$\lg({L_{\rm bol}}/erg/s) \approx 33-34$ with X-ray luminosities varying 
between $\lg{({L_{\rm x}}/erg/s)} \sim $27.5 and 31. 
On the other hand, the early-type primaries show lower 
X-ray luminosities that decrease from  B1 to A0 spectral types. 
However, the three late-B type stars which clearly deviate 
in Fig. 3  also deviate in this figure, 
showing a higher X-ray emission than expected
for their spectral types: B9, B7 and B9.5.

We have compared these results with those previously obtained 
by Schmitt et al. (1993). The  X-ray luminosities 
derived for their sample of 7 Lindroos systems 
agree with our results except in one case: HD\,113703. 
While these authors identify the single HRI X-ray detection with 
the early-type primary, we think it is most probably 
related to the late-type secondary. As we discussed in Sect. 3,
the difference between the X-ray detection and the optical position 
is smaller for the late-type secondary in 
both the broad and hard band images.

At this point we must also remark that the young star HD\,109573\,B 
(HR4796\,B) shows an X-ray luminosity higher than that reported 
by Jura et al. (1998). These authors compute the X-ray luminosity 
from the same HRI image but using an ECF that corresponds
to the PSPC detector, although these detectors have 
different sensitivities. Moreover, they take a mean ECF  
from Neuh\"auser et al. (1995) which is deduced from a ROSAT survey  
on Taurus, a star forming region where most of the stars show 
visual extinctions larger than those in TWA. Therefore, we think that 
the value listed in Table 5 is more realistic since it takes 
into account the ECF from the HRI detector and a negligible 
absorption to the source.

 One of our binary systems, HD\,560, was detected but 
unresolved by the ROSAT HRI (see Fig. 1). 
Because this system is comprised 
of a B9 primary and a G5 type companion, we have made a simple 
test to confirm the results obtained for HD\,123445\,A, HD\,127971\,A 
and HD\,129791\,A: although we certainly know that HD\,560 is 
a binary system, we have supposed HD\,560 to be a single B9 star 
and not a pair. Therefore, if we assume that the single 
X-ray detection corresponds to a B9-type star and we plot this 
source into Fig. 4 (starred symbol), the result is that HD\,560 lies in 
the same part of the diagram as the three sources with
unresolved late-type candidates. Hence, this test strengthens the 
idea of unresolved late-type companions in these late-B type stars.

 We have compared our sample of primary stars with a sample 
of MS B-stars taken from the {\it The RASS catalogue of optically 
bright OB-type stars} (Bergh\"ofer et al., 1996) in Figure 4. 
In principle, our primary stars are 
in good agreement with the sample of B-type stars  with  comparable 
X-ray luminosities. Note, however, the large scatter in
the X-ray values at $\lg{(L_{\rm bol}/erg/s)} \sim$ 35.5 (B9 stars).  
HD\,123445\,A, HD\,127971\,A and HD\,129791\,A
lie in this region of the diagram with X-ray 
luminosities ranging from $10^{29.5}$ to $10^{30.8}$ erg/s. 

 The wide range of X-ray luminosities found in late-B stars from
Bergh\"ofer's  sample  together with the  position of our late-B 
primary stars in Fig. 4, suggests the possibility of having 
unresolved late-type companions in late-B type stars with 
$\lg{(L_{\rm x}/erg/s)}\,>$29.5. Although it is beyond the scope of this paper,
it would be interesting to check if the MS late-B type stars from  
Bergh\"ofer et al. (1996) with highest X-ray luminosities 
are suspected to have late-type unresolved companions.  
 
 As we mentioned before, the $\lg{(L_{\rm x}/L_{\rm bol})}$ ratio is generally 
similar for stars of the same spectral type. As an example, 
a ``canonical'' relation of $(L_{\rm x}/L_{\rm bol}) \approx 10^{-7} $
has been reported for O- and early B-type stars (Harnden et al. 
1979, Long \& White 1980, Pallavicini et al. 1981, Sciortino 
et al. 1990). Given that the $(L_{\rm x}/L_{\rm bol})$ ratio 
is characteristic for each spectral type, we have finally compared
this value with the HRI HR's computed in Sect. 4.2. 

\begin{table*}
\caption{Stellar properties of the PTTS candidates}
\begin{flushleft}
\begin{tabular}{lllllllllll} \hline \noalign{\smallskip}
    HD & {\bf Sp.T.} & {\bf log L$_x$}   & {\bf EW Li I$^1$} & \multicolumn{2}{l} {\bf far-IR data$^2$} 
       &   {\bf RV$^3$} & \multicolumn{4}{l}{\bf Physical companions?}\\
       &  &    & & F(12$\mu$m) & F(100$\mu$m) & & X-rays & Li {\sc I}  & IR-ex. & RV \\   
       & & (erg/s)         & (\AA)    & (Jy)        & (Jy)  &  & & &  & \\ \hline  
   \noalign{\smallskip}
560 B?    & G5 & 30.56      & 0.290      & 0.278     & 1.037 & E & + & + &  + & + \\
17543 C   & F8 & 30.17      & 0.12       & 0.269     & 1.089 &  & + &  + & +  &   - \\ 
27638 B   & G2 & $<$28.49   & 0.152      & 0.375     & 2.014 & E & - &  +  & +  & +  \\ 
33802 B   & G8 & 30.03      & 0.318      & 0.508     & 2.718 & E:    & + & + & + & (+)  \\
38622 B   & G2 & 29.99      & 0.203$\pm$0.055$^{**}$ & 0.263 & 3.056 & E & + & + & +& + \\
40494 B   & G8 &  $<$28.57  & 0.206$\pm$0.013$^*$    & 0.417 & 0.618 &   &  - & +  & +  & ? \\
53191 B   & G3 & 29.53      & 0.222$\pm$0.015$^*$    &       &       & & + & + & ? &?  \\
60102 B   & G8 & 30.62      & 0.252$\pm$0.010$^*$    &       &       & & + & + & ? & ? \\
86388 B ? & F5 & 29.42      & 0.053$\pm$0.004$^*$    &       &       & {\it E} & + & + &   ? &  + \\
87901 B   & K0 & 28.76      & 0.017	             &       &       & E:  & +   & + & ? & (+)  \\ 
90972 B   & F9 & 30.11      & 0.147	             & 0.253 & 1.657 & {\it E:}  & + & + & + & (+) \\
108767 B  & K2 & 28.94      & 0.175	             & 2.563 & 0.497 &  E  & + & + & - & + \\
109573 B  & M2.5 & 29.75    & 0.55$^{***}$           &       &       &     & + & + & ? & ?  \\
113703 B? & K0 & 30.57      & 0.367	             & 0.296 & 1.555 &  {\it E}  & + & + & + & + \\
113791 B  & F7 & 29.98      & 0.136$\pm$0.005$^*$    & 0.452 & 1.581 &  {\it E}  & + & + & + & + \\
123445 & K2 &  $<$29.21 & $\leq$0.04$^*$       &       &       &     & - & - & ? & ? \\
127304 B  & K1 & 28.89      & 0.104$\pm$0.004$^{**}$ & 0.153 & 0.379 &  E  &  + & + & - & + \\
127971 B  & K0 & $<$27.90   & $\le$0.030$^*$         &       &       &     &  - & - & ? & ?  \\
129791 B $\dagger$  & K5 & 30.63      & 0.230        &  ?    & ?     &  & + & + & ? & ? \\
143939 B $\dagger$   & K3 & 30.50      & 0.400   &  ?   &  ?    &  & +  & + & ? & ?  \\  \noalign{\smallskip}\hline
  
 \end{tabular}
 \end{flushleft}
{\bf Notes:} 1. Most of the Li  {\sc I} EW values have been adopted Pallavicini 
et al. (1992). These measurements were derived from high-resolution 
spectra (resolution of 0.1\AA), 
except those marked with an asterisk, which were obtained from low-resolution
spectra (resolution of 2\AA; the data marked with a double asterisk are taken from Mart\'{\i}n 
et al., 1992 (disp. of 0.22\AA/pix); $^{***}$ adopted from Webb et al. (1999); 2. The IR data 
are taken from Ray et al. (1995); the two sources marked with a $\dagger$ 
are located close to the Galactic plane and this seems to be the reason
of not being detected by IRAS; 3. If similar radial velocities have been 
measured for both members of the pair it is indicated by an 
'{\it E}' (if taken from Gahm et al. 1983)
and by an 'E' (if taken from Mart\'{\i}n et al. 1992). 
If the measurements are doubtful,
it is indicated by a ':' (see Gahm et al., 1982 and 
Martin et al., 1992 for more details);
4. We have summarized the data
of the table in this column using the symbols +, - and ? for
positive, negative and unknown properties respectively. The (+)
symbols indicate doubtful properties.

\end{table*}

   \begin{figure*}
   \vspace{-0.2cm}
   \resizebox{18cm}{!}{\includegraphics{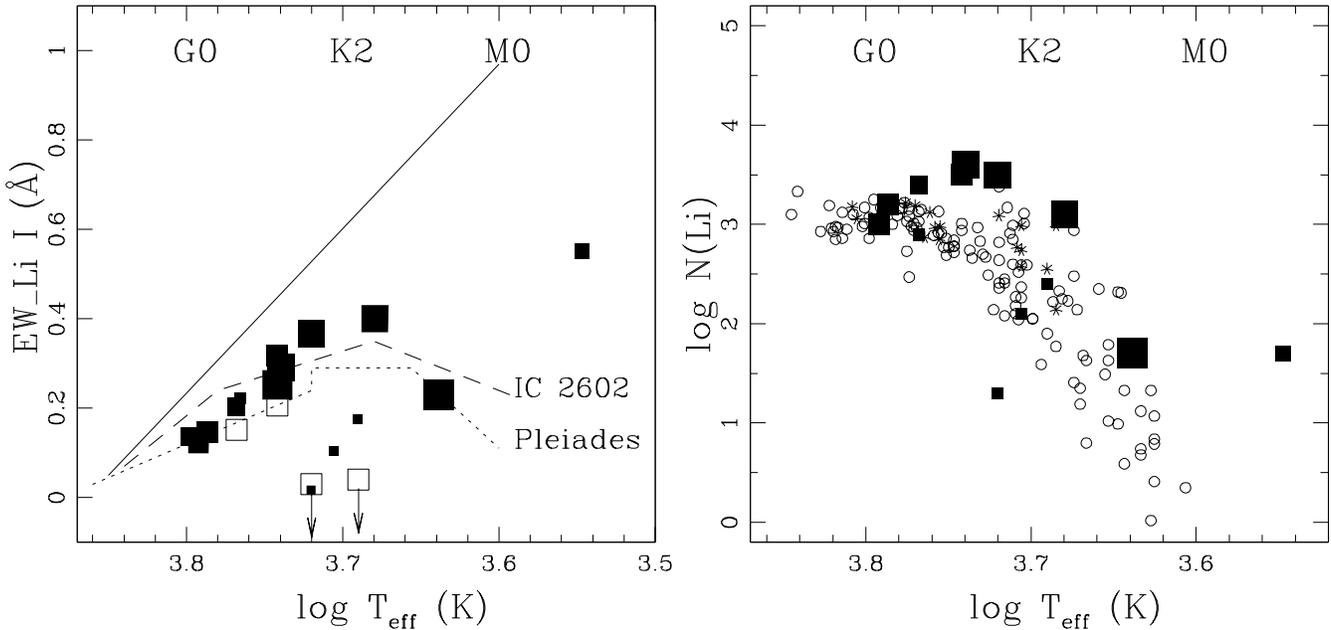}}
     \vspace{-9.5cm}
     \caption{{\bf a} Lithium {\sc I} equivalent width (EW) versus the log
      of the effective temperature for the sample of Lindroos secondaries. 
      The stars are represented by solid squares which size is
      proportional to their X-ray luminosity.
      The open squares show those stars with X-ray upper limits.
      We have also plotted the upper envelopes to the Li {\sc I} EW of the 
      Pleiades (dotted line) and IC\,2602 (dashed line) stars 
      as well as the primordial Li abundance (solid line). {\bf b} Li {\sc I} 
      abundances versus the log of the effective temperature.
      The Lindroos sample is represented by solid squares. The
      Pleiades and IC\,2602 data are represented by open circles
      and starred symbols, respectively. A general trend is seen: most of the 
      late-type stars under study have more Lithium {\sc I} 
      than ZAMS cluster stars, ie. are younger, and the larger the X-ray 
      luminosity the more Lithium abundance at any given spectral type.}
     \label{litemp}  
    \end{figure*}

Fig. 5 shows a clear separation between the stars
of our sample. A boundary line drawn at  $HR=0$ 
allows us to classify our stars in soft and hard
X-ray emitters. While the softer X-ray emitters
lie close to the $\lg{(L_{\rm x}/L_{\rm bol})}\sim 7$ ratio reported 
for  O- and early-B type stars, those with positive HR, 
are spread over a wide range of $\lg{(L_{\rm x}/L_{\rm bol})}$ values. 
Among the latter, we can also distinguish two groups. 
On one hand, we find the late-type secondaries
lying at the top right corner of the figure. On the other hand,
the three late-B type stars with possible unresolved late-type
companions occupy a band with $\lg{(L_{\rm x}/L_{\rm bol})}$ ranging 
from -4.5 to -6. Note that these stars clearly deviate from the group 
of early-type primaries at the left bottom corner 
of the figure. 

 If we include HD\,560 in Fig. 5 making the same assumptions 
as in Fig. 4, ie. assuming that it is a single B9 star responsible
for the detected X-ray emission, we can see that the source also 
lies in the same region as HD\,123445\,A, HD\,127971\,A 
and HD\,129791\,A.

 As a conclusion, we can confirm that B-type stars in the Lindroos
systems under study generally show a decrease in the X-ray 
luminosity for decreasing spectral types (from B0 to B9). 
The $(L_{\rm x}/L_{\rm bol})$ ratio is in agreement with their spectral types 
and it is well-correlated with their negative HR's. However, there are 
three sources, HD\,123445\,A, HD\,127971\,A and HD\,129791\,A which show
$L_{\rm x}$ values higher than those reported for earlier 
B-type stars. Moreover, when comparing the $(L_{\rm x}/L_{\rm bol})$ ratio 
with the computed HRI HR's, these three late-B type stars clearly deviate
from the sample of primary stars, showing values
closer to those reported for the late-type secondaries.  
Therefore, these three late-B  type stars are suspected to have 
unresolved late-type companions.

\section{X-ray emission from late-type secondaries: are they Post 
T Tauri stars?}

As mentioned in the introduction, the late-type 
secondaries of Lindroos systems have been studied in several 
spectral ranges. In particular, the optical and IR data
provide clear evidences of youth among these stars.

In Table 7 we ahow some of the main 
observational properties of these late-type secondaries.
Columns 1, 2 and 3 provide the name of the source, 
its spectral type and the X-ray luminosity derived from this work. 
Column 4 shows the equivalent width (EW) of
the Li {\sc I} absorption line while IR data from the {\it IRAS} 
satellite are provided in columns 5 and 6. 
A `flag' related to the measured radial velocity
of the pair is given in column 7.  We have finally summarized 
all these data in the last columns of the table in order to
isolate those late-type stars with evidences 
to be physically bound to the early-type primaries. 
Apart from the measured radial velocity
of the pair and the Li I EW  we have 
also considered other indicators of youth like the X-ray 
emission and the measured IR excesses. 

 As we can see from Table 7, all the late-type secondaries
in binary systems with similar radial velocities 
show indications of youth. 
For the stars with no radial velocity measurements, we have found pairs 
with clear evidences of youth and hence, 
most probably physically bound to  
their primaries: HD\,17543, HD\,53191, HD\,60102, HD\,129791 and HD\,143939.
HD\,123445\,B and HD\,127971\,B do not seem to be  bound to their 
primaries because they lack indicators of youth.
Finally, HD\,40494\,B  is not detected
in X-rays but it shows a strong Li I absorption line (as HD\,27638). 
Therefore, it is not obvious how to classify this source.
Radial velocity measurements would be convenient to confirm 
the nature of this pair.

According  to Lindroos (1985), most of these stars show ages lower 
than 70 Myr but, as discussed in L86, the uncertainties associated 
with the age 
determination are large. Note that the ages of the Lindroos systems 
were first determined from uvby$\beta$ photometry of the primaries 
(Lindroos 1985) and making use of the evolutionary tracks and isochrones 
by Hejlesen (1980). A new estimation of the ages using modern isochrones 
plus new optical 
and IR photometrical data will be postponed to a later publication.

 Instead of studying the ages of the systems we have considered the 
Li {\sc I} (6708\AA) EW and we have related it to the derived X-ray 
luminosities.  As shown in Table 7, the Li {\sc I} EW has been directly 
measured in most of the Lindroos secondaries, so we can study if there 
is a relation between both parameters. We have plotted in Fig. 6 
the  Li {\sc I} EW versus the effective temperature of our stars. 
Note that these effective temperatures have been derived from their 
spectral types according to the conversion given by Kenyon \& Hartmann (1995). 
We have also considered the upper envelopes to the Li {\sc I } EW 
of two young clusters, the Pleiades ($\sim$ 125 Myr) and IC\,2602 
($\sim$ 35 Myr), so that those secondary stars with Li {\sc I} EW 
stronger than in the Pleiades or 
in IC\,2602 are younger than these clusters (data from 
Soderblom et al. 1993, Randich et al. 1997 and Stauffer et al. 1997).

 Fig. 6a shows that most of the  strongest X-ray emitters are located 
above the Pleiades upper envelopes to the Li {\sc I}. This suggests that 
a large fraction of the Lindroos secondaries are younger than 125 Myr, 
with 4 of them showing ages lower than 35 Myr. Among these 4 stars, 
we can clearly distinguish HD\,109573\,B (HR\,4796\,B) 
in the right middle part of the figure, with a high value of 
the Li {\sc I} EW. Its X-ray emission is not as large as this coming from
the other three sources above the IC2602 Li I envelope. The different
spectral type as well as the different evolutionary state could explain 
this difference. 

Fig. 6a also allows us to confirm the nature
of the HD\,123445 and HD\,127971 pairs. The secondary 
stars lie at the bottom of the diagram with upper 
limits in both, the Li I EW and the X-ray emission. 
Two more stars show non-detections
in X-rays although they lie very close to the upper envelope 
of the Pleiades: HD\,27638 and HD\,40494. As we have discussed 
above, the former is probably bound to the primary 
(see table 7) while the nature of the later one is not still clear. 

As a conclusion, we can say that 
those stars with strong X-ray emission seem to be younger than IC\,2602.
Note, however that although the uncertainties 
associated with the Li {\sc I} EW measurements are
generally low (see Table 7), for the stars earlier 
than $\sim$K0 the Li {\sc I} EW does not allow us to reach 
any conclusion about their ages because F- and G-type PMS 
together with ZAMS stars 
still have their initial Li content.

To study the last issue in detail, we have converted the Li {\sc I} EW into 
lithium abundances, N(Li) making use of non-LTE curves of growth from 
Pavlenko \& Magazz\'u (1996) and assuming a
surface gravity of $\lg{g_s} = 4.5$.   Note that we have only 
considered high-resolution data to make this conversion (see Table 7). 
We have plotted this value against the  effective temperature of our stars 
in Fig. 6b. As in the left panel of the figure, we have found strong
X-ray emitters above the Pleiades and IC\,2602 data.
Note that most of the Lindroos secondaries  (11) show  lithium 
abundances close to 3.0 which is consistent with low mass PMS stars.  

\section{Results and Conclusions}

We have reported the X-ray emission from Lindroos binary systems
observed with ROSAT. Most of the stars from this sample are 
detected in the RASS and several binary systems were observed with the 
PSPC and the HRI. After the analysis of the data, 
the main conclusions are:

{\bf 1.} Both, early- and late-type stars, show the same
distribution of the X-ray Luminosity Function (XLF).
Both samples are statistically similar with a probability of a 0.8. 
When we repeat the analysis without including 
stars that could be contaminating the samples, that is,
without considering most certainly unbound secondaries, early-type primaries
with possible unresolved late-type companions and one unreliable
X-ray detection, the result does not change. The probability is reduced 
to 0.4 but it is large enough to statistically confirm the similarity
of both samples.

{\bf 2.} A careful study of the HRI hardness ratios of our
sample allows to discriminate between soft and hard X-ray
emitters. Late-type stars always show positive HR while most of the
early-type stars show negative values.  Moreover, if we compare the
X-ray luminosity in the two energy bands of the HRI, we can see that
there is a clear separation between the two samples.  Late-type stars
show higher H-band than S-band X-ray luminosities. This
result can be explained in terms of an energetic corona and
activity episodes commonly reported in these stars. In the case of
early-type stars, most of them present higher S-band than H-band
luminosities.  However, there are three X-ray sources 
formally identified as early-type primaries which show 
H-band luminosities comparable to those of late-type 
stars: HD\,123445\,A, HD\,127971\,A and HD\,129791\,A.

{\bf 3.} The $L_{\rm x}\,-\,L_{\rm bol}$ diagram 
shows a clear separation among sources of different spectral
types. This separation is even more prononced when
the $L_{\rm x}/L_{\rm bol}$ ratio is plotted vs. the HRI hardness ratio. While
early-type stars, which generally show negative HR's, lie close to the
$L_{\rm x}/L_{\rm bol}=\,10^{-7}$ ``canonical'' relation, 
late-type stars show positive HR's and present higher values of 
this ratio ($L_{\rm x}/L_{\rm bol}\sim 10^{-3}$).  

For those particular cases of early-type stars with positive HR's, we have seen
that they display X-ray luminosities comparable to those of the
late-type companions.  When they are plotted into the
$L_{\rm x}/L_{\rm bol}$ - HRI HR diagram, they lie closer to 
the late-type stars
than to the B-type group.  All these evidences make us conclude that
HD\,123445\,A, HD\,127971\,A and HD\,129791\,A are good candidates to have
unresolved late-type companions. 

{\bf 5.} The computed X-ray luminosities together 
with the Li I (6708\AA) EW and abundances deduced for the 
Lindroos late-type secondaries have revealed this group of stars
to be a good sample of PTTS candidates. A strong X-ray emission
is reported for the youngest ones. Although we have no 
reliable measurements of their ages, most of the Lindroos
secondaries seem to be younger than 125 Myr (when compared to the Pleiades)
with three of them showing ages lower than 35 Myr (IC\,2602 cluster). 
Also HD\,109573\,B (HR\,4796\,B) is located above these two 
clusters showing a high X-ray emission.

This study also has allowed us to confirm the optical nature of 
the HD\,123445 and HD\,127971 pairs. In this sense, two more systems are
classified as doubtful: HD\,87901 and HD\,40494. The former 
lacks clear indicators of youth while HD\,40494, although not
detected in X-rays, may be a young star bound to its primary.

\begin{acknowledgements}

We would like to thank T. Hearty, M. Fern\'andez, J. Alves, 
S. D\"obereiner and B. K\"onig for their assistance 
and their useful comments. We are very grateful to T. Bergh\"ofer 
for providing his data. Some archived ROSAT observations investigated here
were performed by T. Bergh\"ofer, J. Krautter, J. Puls, T. Simon and J. 
Staubert as PI's. The ROSAT project is supported by the 
Max-Planck-Society and the German Government (DLR/BMBF).

\end{acknowledgements}

\end{document}